\documentclass[final,journal]{IEEEtran}
\usepackage{cite}

\ifCLASSINFOpdf
   \usepackage[pdftex]{graphicx}
  \DeclareGraphicsExtensions{.pdf,.jpeg,.png}
\else
   \usepackage[dvips]{graphicx}
   \DeclareGraphicsExtensions{.eps}
\fi
\usepackage[cmex10]{amsmath}
\usepackage{array}
\usepackage{ifpdf,amssymb,amsthm,mathtools,amsmath}

\begin{document}
%
\title{The Optimal Hard Threshold \\for Singular Values is  $4/\sqrt{3}$}
%
%
%

\author{Matan~Gavish~
        and~David~L.~Donoho
\thanks{The authors are with the Department of Statistics, Stanford University,
Stanford, CA 94305 USA (e-mail: gavish@stanford.edu; donoho@stanford.edu)}
}

\theoremstyle{theorem} \newtheorem{thm}{Theorem}
\theoremstyle{definition} \newtheorem{cor}{Correlary}
\theoremstyle{definition} \newtheorem{defn}{Definition}
\theoremstyle{definition} \newtheorem{lemma}{Lemma}
\theoremstyle{definition} \newtheorem{prop}{Proposition}
\newcommand{\R}{\ensuremath{\mathbb{R}}}
\newcommand{\E}{\ensuremath{\mathbb{E}}}
\newcommand{\C}{\ensuremath{\mathbb{C}}}
\newcommand{\Kc}{\mathcal{K}}
\newcommand{\Nc}{\mathcal{N}}
\newcommand{\Lc}{\mathcal{L}}
\newcommand{\Rc}{\mathcal{R}}
\newcommand{\Dc}{\mathcal{D}}
\newcommand{\Cc}{\mathcal{C}}
\newcommand{\Gc}{\mathcal{G}}
\newcommand{\Uc}{\mathcal{U}}
\newcommand{\Mc}{\mathcal{M}}
\newcommand{\Yc}{\mathcal{Y}}
\newcommand{\Xc}{\mathcal{X}}
\newcommand{\Pc}{\mathcal{P}}
\newcommand{\Fc}{\mathcal{F}}
\newcommand{\Ac}{\mathcal{A}}
\newcommand{\V}[1]{\ensuremath{\mathbf{#1}}}
\newcommand{\norm}[1]{\left|\left| #1 \right|\right|}
\newcommand{\bt}[1]{\tilde{\mathbf{#1}}}
\newcommand{\plim}{\stackrel{p}{\rightarrow}}
\newcommand{\aslim}{\stackrel{a.s.}{\longrightarrow}}
\newcommand{\aseq}{\stackrel{a.s.}{=}}
\newcommand{\dlim}{\stackrel{d}{\rightarrow}}
\newcommand{\iid}{\stackrel{\text{iid}}{\sim}}
\newcommand{\ind}{\stackrel{\text{ind.}}{\sim}}
\newcommand{\eqdef}{\stackrel{\text{def}}{=}}
\newcommand{\indep}{\stackrel{\text{indep.}}{\sim}}
\newcommand{\minimize}[1]{\text{minimize }\,& #1}
\newcommand{\st}[1]{\text{subject to }\,& #1}
\newcommand{\Pd}[3]{\ifthenelse{\equal{#3}{1}}{\frac{\partial #1}{\partial #2}}{\frac{\partial^{#3} #1}{\partial #2^{#3}}}}
\newcommand{\inv}[1]{\frac{1}{#1}}
\newcommand{\Mn}{M_{\n}}
\newcommand{\Mnr}{M_{\n}(r)}
\newcommand{\Mnnr}{M_{n\times n}(r)}
\newcommand{\Fce}{\Pc_{n,\varepsilon}}
\newcommand{\Pce}{\Pc_{n,\varepsilon}}
\newcommand{\sv}{\Sigma}
\newcommand{\on}{\Delta}
\newcommand{\off}{ { \Delta^{C} }  }
\newcommand{\lf}{LF}
\newcommand{\n}{n} 
\newcommand{\m}{m} 
\newcommand{\rk}{r} 
\newcommand{\Mnn}{M_{\n\times\n}}
\newcommand{\Mmn}{M_{\m\times\n}}
\newcommand{\Mmm}{M_{\m\times\m}}
\newcommand{\MSE}{\mathbf{M}}
\newcommand{\mmx}{\mathcal{M}}
\newcommand{\X}{\mathbf{X}}
\newcommand{\Xmn}{\mathbf{X}_{\m,\n}}
\newcommand{\Xnn}{\mathbf{X}_{\n,\n}}
\newcommand{\z}{\zeta}
\newcommand{\cM}{\mmx}
\newcommand{\bR}{{\bf R}}
\newcommand{\bC}{{\bf C}}
\newcommand{\bZ}{{\bf Z}}
\newcommand{\bX}{{\bf X}}
\newcommand{\bI}{{\bf I}}
\newcommand{\cA}{{\cal A}}
\newcommand{\cL}{{\cal L}}
\newcommand{\bitem}{\begin{itemize}}

\newcommand{\eitem}{\end{itemize}}
\newcommand{\goto}{\rightarrow}
\newcommand{\beq}{\begin{equation}}
\newcommand{\eeq}{\end{equation}}
\newcommand{\bx}{\V{x}}
\newcommand{\by}{\V{y}}
\newcommand{\bz}{\V{z}}
\newcommand{\eps}{\varepsilon}
\newcommand{\sgn}{\text{sign}}
\newcommand{\inner}[2]{\ensuremath{\langle #1\,,\,#2 \rangle}}


\maketitle

\begin{abstract}
   
  We consider recovery of low-rank matrices from noisy data by hard thresholding
  of singular values, in which empirical singular values below a threshold
  $\lambda$ are set to $0$. We study the asymptotic MSE (AMSE) in a framework
  where the matrix size is large compared to the rank of the matrix to be
  recovered, and the signal-to-noise ratio of the low-rank piece stays constant.
  The AMSE-optimal choice of hard threshold, in the case of $n$-by-$n$ matrix in
  white noise of level $\sigma$, is simply $(4/\sqrt{3}) \sqrt{n}\sigma  \approx
  2.309 \sqrt{n}\sigma$ when $\sigma$ is known, or simply $2.858\cdot y_{med}$
  when $\sigma$ is unknown, where $y_{med}$ is the median empirical singular
  value. For nonsquare $m$ by $n$ matrices with $m \neq n$ the thresholding
  coefficients $4/\sqrt{3}$ and $2.858$ are replaced with different provided
  constants that depend on $m/n$.  Asymptotically, this thresholding rule adapts
  to unknown rank and unknown noise level in an optimal manner: it is {\em
  always} better than hard thresholding at any other value, and is {\em always}
  better than ideal Truncated SVD (TSVD), which truncates at the true rank of
  the low-rank matrix we are trying to recover.  Hard thresholding at the
  recommended value to recover an $n$-by-$n$ matrix of rank $\rk$ guarantees an
  AMSE at most $3\,\n\rk \sigma^2$.  In comparison, the guarantees provided by
  TSVD,  optimally tuned singular value soft thresholding and the best guarantee
  achievable by any shrinkage of the data singular values are
  $5\,\n\rk\sigma^2$, $6\,\n\rk\sigma^2$, and $2\,\n\rk\sigma^2$, respectively.
  The recommended value for hard threshold also offers, among hard thresholds,
  the best possible AMSE guarantees for recovering matrices with bounded nuclear
  norm. Empirical evidence suggests that performance improvement over TSVD and
  other popular shrinkage rules can be substantial, for different noise
  distributions, even in relatively small $n$.
 
\end{abstract}


%
\IEEEpeerreviewmaketitle

\section{Introduction} 

\IEEEPARstart{S}{uppose} that we are interested in an
unknown $m$-by-$n$ matrix $X$, thought to be either exactly or approximately of
low rank, but we only observe a single noisy $\m$-by-$\n$ matrix $Y$, obeying
$Y=X+\sigma Z$. The noise matrix $Z$ has independent, identically distributed,
zero-mean entries.  The matrix $X$ is a (non-random) parameter, and we wish to
estimate  it with some bound on the mean squared error (MSE). 

The default estimation technique for our task is {\em Truncated SVD} (TSVD)
\cite{Golub1965}: Write
\begin{eqnarray} \label{svd:eq}
  Y = \sum_{i=1}^\m y_i \V{u}_i \V{v}_i'
\end{eqnarray}
for the Singular Value Decomposition of the data matrix $Y$, where
$\V{u}_i\in\R^m$ and $\V{v}_i\in\R^n$, $i=1,\ldots,m$ are the left and right
singular vectors of $Y$ corresponding to the singular value $y_i$.
The TSVD estimator is 
\begin{eqnarray*}
  \hat{X}_\rk = \sum_{i=1}^\rk y_i \V{u}_i \V{v}_i'\,,
\end{eqnarray*}
where $\rk=rank(X)$, assumed known, and $y_1\geq \ldots \geq y_m$.  Being the
best approximation of rank $\rk$ to the data in the least squares sense
\cite{Eckart1936}, and therefore the Maximum Likelihood estimator when $Z$ has
Gaussian entries, the TSVD is arguably as ubiquitous in science and engineering
as linear regression
\cite{Alter2000,Cattell1966,Jackson1993,Lagerlund1997,Price2006,Edfors1998}.

When the true rank $r$ of the signal $X$ is unknown, one might try to form an
estimate $\hat{r}$ and then apply the TSVD $\hat{X}_{\hat{r}}$.  Extensive
literature has formed on methods to estimate $r$: we point to the early
\cite{Cattell1966,Wold1978} (in Factor Analysis and Principal Component
Analysis), the recent \cite{Hoff2006,Owen2009,Perry2009} (in our setting of
Singular Value Decomposition), and reference therein. 
It is instructive to think about rank estimation (using any method), followed by
TSVD, simply as {\em hard thresholding} of the data singular values, where only
components $y_i\V{u}_i\V{v}_i'$ for which $y_i$ passes a specified threshold,
are included in $\hat{X}$.  Let $\eta_H(y,\tau) = y \mathbf{1}_{\{y \geq \tau
\}}$ denote the hard thresholding nonlinearity, and consider Singular Value Hard
Thresholding (SVHT) \begin{eqnarray}\label{hard:eq} \hat{X}_\tau = \sum_{i=1}^\m
  \eta_H(y_i;\tau)\, \V{u}_i \V{v}_i'\,.  \end{eqnarray} In words,
$\hat{X}_\tau$ sets to $0$ any data singular value below $\tau$.

\begin{figure}[h] 
  \begin{center} 
    \includegraphics[width=3.7in]{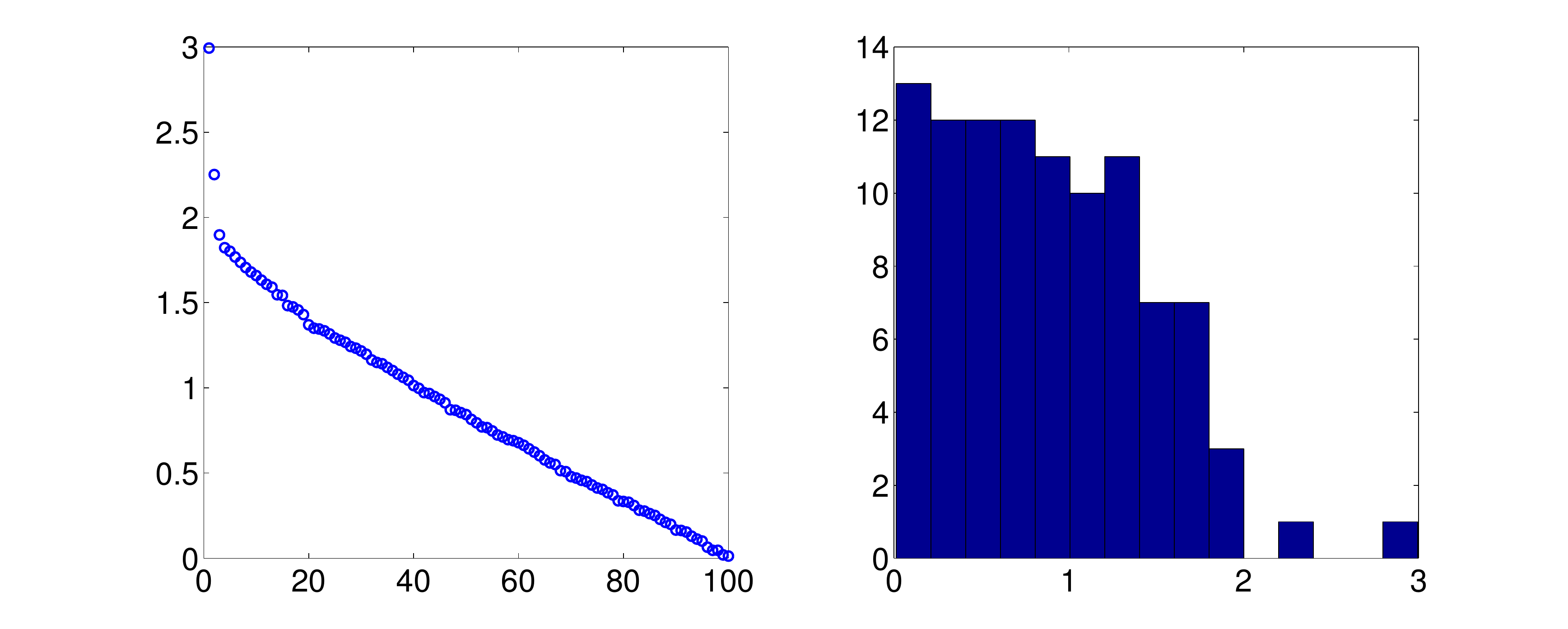}
  \end{center} 
  \caption{\scriptsize Singular values of a data matrix $Y\in
    M_{100,100}$ sampled from the model $Y=X+Z/\sqrt{100}$, where $X_{1,1} =
    1.7$, $X_{2,2}=2.5$ and $X_{i,j}=0$ elsewhere. In Matlab, a sample can be
    generated using the command {\tt y=svd(diag([[1.7 2.5]
    zeros(1,98)])+randn(100)/sqrt(100))}.  Left: the singular values of $Y$
    plotted in decreasing order, known in Principal Components Analysis as a
    Scree plot. Right: the singular values of $Y$ in a histogram with $15$ bins.
    Note the bulk edge approximately at $2$, and note that the location of the
    top two singular values of $Y$ is approximately $X_{i,i} + 1/X_{i,i}$
    ($i=1,2$), in agreement with Lemma \ref{y-asy:lem}. 
  } 
    \label{hist:fig}
  \end{figure}

Matrix denoisers explicitly or implicitly based on hard thresholding of singular
values have been proposed by many authors, including
\cite{Achlioptas2001,Azar2001,Bickel2008,Chatterjee2010,
Keshavan2009,Keshavan2010,Owen2009,Perry2009,Tanner2012}.  As a common example
of implicit SVHT denoising, consider the standard practice of estimating $r$ by
plotting the singular values of $Y$ in decreasing order, and looking for a
``large gap'' or ``elbow'' (Figure \ref{hist:fig}, left panel).  When $X$ is
exactly or approximately low-rank and the entries of $Z$ are white noise of zero
mean and unit variance, the empirical distribution of the singular values of the
$m$-by-$n$ matrix $Y=X+\sigma Z$ forms a quarter-circle bulk whose edge lies
approximately at $(1+\sqrt{\beta})\cdot \sqrt{n}\sigma$, with $\beta=m/n$
\cite{silverstein_book}.  Only data singular values that are larger than the
bulk edge are noticeable in the empirical distribution (Figure \ref{hist:fig},
right plot).  Since the singular value plot ``elbow'' is located at the bulk
edge, the popular method of TSVD at the ``elbow'' is an approximation of {\em
bulk-edge hard thresholding}, $\hat{X}_{(1+\sqrt{\beta})\sqrt{n}\sigma}$.

\subsection{Questions}

Let us measure the denoising performance of a denoiser $\hat{X}$ at a
signal matrix $X$ using Mean Square Error (MSE), 
\begin{eqnarray*}
  \norm{\hat{X}(Y)-X}_F^2=\sum_{i,j}(\hat{X}(Y)_{i,j}-X_{i,j})^2\,.
\end{eqnarray*}
The TSVD is an optimal rank-$r$ approximation of the data matrix $Y$, in MSE.
But this does not necessarily mean that it is a good, or even reasonable,
estimator to the signal matrix $X$, which we wish to recover.  We may wonder:
\begin{itemize}
  \item {\em Question 1.} 
    Assume that $rank(X)$ is unknown but small. Is there a singular value
    threshold $\tau$ so that SVHT $\hat{X}_\tau$ successfully adapts to unknown
    rank and unknown noise level, and performs as well as TSVD would, had we
    known the true $rank(X)$?
\end{itemize}

As we will see, it is convenient to represent the threshold as $\tau = \lambda
\sqrt{n} \sigma$, where $\lambda$ is a parameter typically between $1$ and $10$.
Recently, S. Chatterjee \cite{Chatterjee2010} proposed that one could have
a single universal choice of $\lambda$; that in a setting more general, but
similar, to our setting, any $\lambda>2$
would give near-optimal MSE, in a qualitative sense; and he specifically
proposed $\lambda=2.02$, namely $\hat{X}_{2.02\,\sqrt{n}\sigma}$ as a
{universal} choice for SVHT, regardless of the shape $\m/\n$ of the matrix, and
regardless of the underlying signal matrix $X$ or its rank. While the rule of
\cite{Chatterjee2010}  was originally intended to be `fairly good' across many
situations not reducible to the low-rank matrix in i.i.d noise model considered
here, $\hat{X}_{2.02\sqrt{n}\sigma}$ is a specific proposal, which prompts the
following question:
\begin{itemize}
  \item {\em Question 2.} 
    Is there really a single threshold parameter $\lambda_*$ that provides 
    good performance guarantees for MSE? Is that value  $2.02$? Is it really
    independent of $m$ and $n$?
\end{itemize}

Finally, note that  singular value hard thresholding is just one strategy for
matrix denoising. It is not a-priori clear whether the whole idea of only
`keeping' or `killing' empirical singular values based on their size makes
sense. Could there exist a shrinkage rule $\eta:[0,\infty)\to[0,\infty)$, that
  more smoothly transitions from `killing' to `keeping', which leads to a much
  better denoising scheme?  We may wonder:
\begin{itemize}
  \item {\em Question 3.} How does optimally tuned SVHT 
    compare with the performance of the 
    {\em best possible shrinkage} of singular values, at least in
the worst-case MSE sense?
\end{itemize}

\subsection{Optimal location for hard thresholding of singular values}

Our main results imply that, in a certain asymptotic framework, there are
simple and convincing answers to these questions.  Following Perry
\cite{Perry2009} and Shabalin and Nobel
\cite{Shabalin2010}, we adopt an asymptotic framework where the matrix grows
while keeping the nonzero singular values of $X$ fixed, and the signal-to-noise
ratio of those singular values stays constant with increasing $n$.  

In this asymptotic framework, 
for a low-rank $n$-by-$n$ matrix observed in white 
noise of level $\sigma$, 
\begin{eqnarray*}
  \tau_*=\frac{4}{\sqrt{3}}\sqrt{n}\sigma\approx 2.309\sqrt{n}\sigma
\end{eqnarray*}
is the optimal location for the hard thresholding of singular values. 
For a non-square $m$-by-$n$ matrix with $m\neq n$, the optimal location is
\begin{eqnarray} \label{taustar-nonsq:eq}
  \tau_* = \lambda_*(\beta) \cdot\sqrt{n}\sigma,
\end{eqnarray}
where $\beta=m/n$.
The value $\lambda_*(\beta)$ is the {\em optimal hard threshold coefficient} for
known $\sigma$. It is given by  formula  
 \eqref{lambdastar-nonsquare:eq} below and tabulated for convenience in Table
 \ref{lambdastar:tab}. 
 (Note added in proof: we found that P. Perry's PhD thesis \cite{Perry2009}
 proposes a threshold which can be shown to be equivalent to
 \eqref{taustar-nonsq:eq}.)

\subsection{Answers}

Our central observation is as follows.  
\begin{quote} {\em When a data singular
    value $y_i$ is too small, then its associated singular vectors
    $\V{u}_i,\V{v}_i$ are so noisy that the component $y_i \V{u}_i \V{v}'_i$
    should not included in $\hat{X}$.  In our asymptotic framework, which models
    large, low-rank matrices observed in white noise, the cutoff below which
    $y_i$ is {\em too small} is exactly $(4/\sqrt{3})\sqrt{n}\sigma$ (for square
  matrices).} 
\end{quote}

\begin{itemize}

  \item {\em Answer to Question 1: Optimal SVHT dominates TSVD.} Optimally tuned
    SVHT $\hat{X}_{\tau_*}$ is {\em always} at least as good as TSVD
    $\hat{X}_\rk$, in terms of AMSE (Theorem \ref{inadmis-tsvd:thm}).  Unlike
    $\hat{X}_\rk$, the optimal SVHT $\hat{X}_{\tau_*}$ does not require
    knowledge of $\rk=rank(X)$. In other words, it adapts to unknown low rank
    while giving uniformly equal or better performance.  For square matrices,
    the TSVD provides a guarantee on worst-case AMSE that is $5/3$ times the
    guarantee provided by $\hat{X}_{\tau_*}$ (Table
    \ref{mmx-square-boundedrank:tab}).
  
  \item {\em Answer to Question 2: Optimal SVHT dominates every other choice of
    Hard Threshold.} In terms of AMSE, optimally tuned SVHT $\hat{X}_{\tau_*}$
    is {\em always} at least as good as SVHT $\hat{X}_\tau$ at any other fixed
    threshold $\tau = \lambda \sqrt{n}\sigma$ (Theorem \ref{inadmis:thm}).  It
    is the asymptotically minimax SVHT denoiser, over matrices of small bounded
    rank (Theorems \ref{mmx-square:thm} and \ref{mmx-nonsquare:thm}) and over
    matrices of small bounded nuclear norm (Theorem \ref{mmx-l1:thm}). In
    particular, the parameter $\lambda = 2.02$ is noticeably worse.  For square
    matrices, $\hat{X}_{2.02\sqrt{n}\sigma}$ provides a guarantee for worst-case
    AMSE that is $4.26/3\approx 1.4$ times the guarantee provided by
    $\hat{X}_{\tau_*}$ (Table \ref{mmx-square-boundedrank:tab}).
  
  \item {\em Answer to Question 3. Optimal SVHT compares adequately to the
    optimal shrinker.} Optimally tuned SVHT $\hat{X}_{\tau_*}$ provides a
    guarantee on worst-case asymptotic MSE that is $3/2$ times (for square
    matrices) the best possible guarantee achievable by {\it any} shrinkage of
    data singular values (Table \ref{mmx-square-boundedrank:tab}).

\end{itemize}

These are all rigorous results, within a specific asymptotic framework, which
prescribes a certain scaling of the noise level, the matrix size, and the
signal-to-noise ratio as $n$ grows.  But  does  AMSE predict actual MSE in
finite-sized problems?  In Section \ref{MSEvsAMSE:subsec} we show finite-$n$
simulations  demonstrating the effectiveness of these results even at rather
small problem sizes.  In high signal-to-noise, all denoisers considered here
perform roughly the same, and in particular the classical TSVD is a valid choice
in that regime.  However, in low and moderate SNR, the performance gain of
optimally tuned SVHT is substantial, and can offer $30\%-80\%$ decrease in AMSE.

\subsection{Optimal singular value hard thresholding -- In practice}

For a low-rank $n$-by-$n$ matrix observed in white 
noise of {\em unknown} level, one can use the data to obtain an approximation of the
optimal location $\tau_*$. Define
\begin{eqnarray*}
  \hat{\tau}_*  \approx
  2.858\cdot y_{med}\,,
\end{eqnarray*}
where $y_{med}$ is the median singular value of the data matrix $Y$.
 The notation $\hat{\tau}_*$ is meant to emphasize
 that this is not a fixed threshold chosen a-priori, but rather a data-dependent 
 threshold.
 For a non-square $m$-by-$n$ matrix with $m\neq n$, 
the approximate optimal location when $\sigma$ is unknown is 
\begin{eqnarray} \label{tauhatstar-nonsq:eq}
  \hat{\tau}_* = \omega(\beta)\cdot y_{med}\,.
\end{eqnarray}
The optimal hard threshold coefficient for
unknown $\sigma$, denoted by $\omega(\beta)$, 
is not available as an analytic formula, but can easily be
evaluated numerically. We provide a
Matlab script for this purpose
\cite{code}; the underlying derivation appears in Section
\ref{sigmahat:subsec} below.
Some values of $\omega(\beta)$ are provided in Table
\ref{lambdastar-hat:tab}. 
When a high-precision value of $\omega(\beta)$ cannot be computed, 
one can use the approximation 
\begin{eqnarray} \label{omega-approx:eq}
\omega(\beta) \approx 0.56\beta^3 - 0.95\beta^2 + 1.82\beta + 1.43\,.
\end{eqnarray}
The optimal SVHT for unknown
noise level, $\hat{X}_{\hat{\tau}_*}$, is very simple to implement and does not require
any tuning parameters. The denoised matrix  
$\hat{X}_{\hat{\tau}_*}(Y)$ can be computed using just a few code lines in a
high-level scripting language. For example, in Matlab:
\begin{verbatim}
   beta = size(Y,1) / size(Y,2);
   omega = 0.56*beta^3 - 0.95*beta^2 + ...
        1.82*beta + 1.43;
   [U D V] = svd(Y);
   y = diag(Y);
   y( y < (omega * median(y) ) = 0;
   Xhat = U * diag(y) * V';
\end{verbatim}
Here we have used the approximation \eqref{omega-approx:eq}. We recommend,
whenever possible, to use a function {\tt omega(beta)}, such as the one we
provide in the code supplement \cite{code}, to compute the coefficient
$\omega(\beta)$ to high precision.

In our asymptotic framework, $\tau_*$ and $\hat{\tau}_*$ enjoy exactly the same
optimality properties. This means that $\hat{X}_{\hat{\tau}_*}$ adapts to
unknown low rank {\em and} to unknown noise level.  Empirical evidence suggest
that their performance for finite $n$ is similar. As a result, the answers we
provide above hold for the threshold $\hat{\tau}_*$ when the noise level is
unknown, just as they hold for the threshold $\tau_*$ when the noise level is
known.

\section{Preliminaries and setting}

Column vectors are denoted by boldface lowercase letters, such as $\V{v}$, their
transpose is $\V{v}'$ and their $i$-th coordinate is $v_i$. The Euclidean inner
product and norm on vectors are denoted by $\inner{\V{u}}{\V{v}}$ and
$\norm{\V{u}}_2$, respectively.  Matrices are denotes by uppercase letters, such
as $X$, its transpose is $X'$ and their $i,j$-th entry is $A_{i,j}$.  $\Mmn$
denotes the space of real $m$-by-$n$ matrices,
$\inner{X}{Y}=\sum_{i,j}X_{i,j}Y_{i,j}$ denotes the Hilbert–-Schmidt inner
product, and $\norm{X}_F$ denotes the corresponding Frobenius norm on $\Mmn$.
For simplicity we only consider $m\leq n$.  We denote matrix denoisers, or
estimators, by $\hat{X}:\Mmn\to\Mmn$.  The symbols $\aslim$ and $\aseq$ denote
almost sure convergence and equality of a.s. limits, respectively.

\subsection{Scaling considerations in singular value thresholding}

With the exception of TSVD, when $\sigma$ is known, all the denoisers we discuss operate by shrinkage of data singular values,
namely are of the form
\begin{eqnarray}
 \label{general-denoiser:eq}
  \hat{X}:  \sum_{i=1}^\m y_i\V{u}_i \V{v}_i' \mapsto
\sum_{i=1}^\m \eta(y_i;\lambda) \V{u}_i \V{v}_i'
 \end{eqnarray}
where $Y$ is given by \eqref{svd:eq} and $\eta:[0,\infty)\to[0,\infty)$ 
is some univariate shrinkage rule. As we will see, 
in the general model $Y=X+\sigma Z$, 
the noise level in the singular values of $Y$ is $\sqrt{n}\sigma$. 
Instead of specifying a different shrinkage rule that depends on the matrix
size $n$, we calibrate our
shrinkage rules to the ``natural'' model $Y=X+Z/\sqrt{n}$. In this convention,
shrinkage rules stay the same for every value of $n$, and we conveniently abuse
notation by writing $\hat{X}$ as in 
\eqref{general-denoiser:eq} for any
$\hat{X}:\Mmn\to\Mmn$, keeping $m$ and $n$ implicit. 
To apply any denoiser $\hat{X}$ below to data from the 
general model $Y=X+\sigma Z$, use the denoiser
\begin{eqnarray} \label{scale-xhat:eq}
  \hat{X}^{(n,\sigma)}(Y) = \sqrt{n}\sigma \cdot \hat{X}(Y/\sqrt{n}\sigma)\,.
\end{eqnarray}
For example, to apply the SVHT
\[\hat{X}_\lambda:\sum_{i=1}^\m y_i \V{u}_i \V{v}_i' \mapsto
\sum_{i=1}^\m \eta_H(y_i;\lambda) \V{u}_i \V{v}_i'\] to data sampled from the model
$Y=X+\sigma Z$, use
$  \hat{X}_{\tau}$, with \[\tau=\lambda\cdot \sqrt{n}\sigma\,.\]
Throughout the text, we use $\hat{X}_\lambda$ to denote SVHT calibrated for
noise level $1/\sqrt{n}$ and $\hat{X}_{\tau}$ to denote SVHT calibrated for a
specific general model $Y=X+\sigma Z$.

To translate the AMSE of any denoiser $\hat{X}$, calibrated for noise level
$1/\sqrt{n}$, to an approximate MSE of the corresponding denoiser $\hat{X}^{(n,\sigma)}$,
calibrated for a model $Y=X+\sigma Z$, we use the identity
\begin{IEEEeqnarray*}{lCl}
  \IEEEeqnarraymulticol{3}{l}{
    \norm{\hat{X}^{(n,\sigma)}(Y)-X}_F^2=}
  \\ \qquad \qquad&&
  n\cdot\sigma^2\cdot\norm{\hat{X}(X/(\sqrt{n}\sigma))+Z/\sqrt{n})-X/(\sqrt{n}\sigma)}_F^2\,.
\end{IEEEeqnarray*}
Below, we spell out this translation of AMSE where appropriate.

\subsection{Asymptotic framework and problem statement} \label{framework:subsec}

In this paper, we consider a sequence of increasingly larger denoising problems $Y_n=X_n+Z_n/\sqrt{n}$, with 
$X_n,Z_n\in M_{m_n,n}$, satisfying the
following assumptions:
\begin{enumerate}
  \item {\em Invariant white noise:} The entries of $Z_n$ are i.i.d
    samples from a
    distribution with zero mean, unit variance and finite fourth moment.  
    To simplify the formal statement of our results, we assume that this
    distribution is {\em orthogonally invariant} in the sense that 
    $Z_n$ follows the same distribution as $A Z_n B$, for every 
    orthogonal $A\in M_{m_n,m_n}$ and $B\in M_{n,n}$. This is the case, for example,
    when the entries of $Z_n$ are Gaussian. 
    In Section \ref{general_noise:sec}   we revisit this restriction and discuss 
    general (not necessarily invariant) white
    noise.
\item {\em Fixed signal column span $(x_1,\ldots,x_r)$:}
Let the rank $\rk>0$ be fixed and choose a vector $\V{x}\in\R^\rk$ with
coordinates $\V{x}=(x_1,\ldots,x_\rk)$ such that $x_1\geq \ldots \geq x_r>0$. Assume that for all $n$, 
\begin{eqnarray} \label{singvec:eq}
X_n = U_n \, diag(x_1,\ldots,x_r,0,\ldots,0) \, V_n'\,
\end{eqnarray}
is an arbitrary\footnote{While the signal rank $r$ and nonzero signal singular values 
$x_1,\ldots,x_r$ are shared by all matrices $X_n$, the signal left and right
singular vectors $U_n$ and $V_n$ are unknown and arbitrary.}
 singular value decomposition of $X_n$, where $U_n\in M_{m_n,m_n}$ and 
$V_n\in M_{n,n}$. 
\item {\em Asymptotic aspect ratio $\beta$:}
    The sequence $\m_\n$ is such that $\m_\n / \n \to \beta$. To simplify our
formulas,
we assume that $0< \beta\leq 1$. 

\end{enumerate}

Let $\hat{X}$ be any singular value shrinkage denoiser calibrated, as discussed above, for noise level $1/\sqrt{n}$.
Define the Asymptotic 
MSE (AMSE) of an $\hat{X}$ at a signal $\V{x}$ by the (almost sure) limit\footnote{
Our results imply that the AMSE is well-defined as a function of the signal
singular values $\V{x}$.
}
\begin{eqnarray} \label{amse_def:eq}
  \MSE(\hat{X},\V{x}) \aseq 
  \lim_{\n\to\infty}
  \norm{\hat{X}(Y_n)-X_\n}_F^2
  \,.
\end{eqnarray}

Adopting the asymptotic framework above, we seek singular value thresholding
rules $\hat{X}_\lambda$ that minimize the AMSE
$\MSE(\hat{X}_\lambda,\V{x}) $.
As we will see, in this framework there are simple, satisfying answers to the
  questions posed in the introduction.

\section{Results}
Define the {\em optimal hard threshold for singular values} for $n$-by-$n$
square matrices by
\begin{eqnarray}
  \lambda_* = \frac{4}{\sqrt{3}}\,.
\end{eqnarray}
More generally, define the optimal threshold for $\m$-by-$\n$ matrices with
$\m/\n=\beta$ by 
\begin{eqnarray} \label{lambdastar-nonsquare:eq}
  \lambda_*(\beta) \eqdef 
  \sqrt{2(\beta+1)+\frac{8\beta}{(\beta+1)+\sqrt{\beta^2+14 \beta+1}}} 
  \,.
\end{eqnarray}
Some values of $\lambda_*(\beta)$ are provided in Table \ref{lambdastar:tab}.
\begin{table}
  \centering
    \begin{tabular}{||r|l||r|l||}
\hline
$\beta$ & $\lambda_*(\beta)$ & $\beta$ & $\lambda_*(\beta)$\\\hline\hline
0.05&1.5066&0.55&2.0167\\\hline
0.10&1.5816&0.60&2.0533\\\hline
0.15&1.6466&0.65&2.0887\\\hline
0.20&1.7048&0.70&2.1229\\\hline
0.25&1.7580&0.75&2.1561\\\hline
0.30&1.8074&0.80&2.1883\\\hline
0.35&1.8537&0.85&2.2197\\\hline
0.40&1.8974&0.90&2.2503\\\hline
0.45&1.9389&0.95&2.2802\\\hline
0.50&1.9786&1.00&2.3094\\\hline
  \end{tabular}
  \caption{\scriptsize Some optimal hard threshold coefficients
  $\lambda_*(\beta)$ from \eqref{taustar-nonsq:eq}.
  For 
  $m$-by-$n$ matrix in known noise level $\sigma$ (with $m/n=\beta$), the 
  optimal SVHT denoiser $\hat{X}_{\tau_*}$ sets to zero 
  all data singular values below the threshold
  $\tau_*=\lambda_*(\beta)\sqrt{n}\sigma$.}
  \label{lambdastar:tab}
\end{table}

\subsection{Optimally tuned SVHT asymptotically dominates TSVD and any SVHT}

Our primary result is simply that $\hat{X}_{\lambda_*}$ always has equal or better AMSE
compared to SVHT with any other choice of threshold, and compared to TSVD. 
In other words, from the ideal perspective of our asymptotic framework, 
the decision-theoretic picture is very
straightforward:
TSVD is asymptotically inadmissible,
and so is
any SVHT with $\lambda\neq\lambda_*$. We note that since AMSE of SVHT with
$\lambda<1+\sqrt{\beta}$ in our framework turns out to be infinite, here and below we need only consider SVHT
with $\lambda > 1+\sqrt{\beta}$. As discussed in Section \ref{conclusion:sec}, 
AMSE calculation in the case where the
threshold $\lambda$ is placed exactly at the bulk edge $1+\sqrt{\beta}$ is a
little more subtle and lies beyond our current scope.

\begin{thm} \label{inadmis:thm}
  {\bf The threshold $\lambda_*$ is asymptotically optimal for SVHT.
  } 
  Let $0<\beta\leq 1$. For any $\lambda>1+\sqrt{\beta}$,  any $\rk\in\mathbb{N}$ and any
  $\V{x}\in\R^\rk$, 
the AMSE \eqref{amse_def:eq} of the SVHT
  denoiser
  $\hat{X}_\lambda$ is well defined and
  \begin{eqnarray}
    \MSE(\hat{X}_{\lambda_*},\V{x}) \leq \MSE(\hat{X}_\lambda,\V{x})\,,
  \end{eqnarray}
  where $\lambda*=\lambda_*(\beta)$ is the optimal threshold
  \eqref{lambdastar-nonsquare:eq}. Moreover, if $\lambda\neq \lambda_*(\beta)$,  
   strict inequality holds at least at one point $\V{x}_*(\lambda)\in\R^r$.
\end{thm}

We can therefore say that $\lambda_*(\beta)$ is {\em asymptotically unique admissible} for
SVHT. In particular, the popular practice of hard thresholding close to the
bulk edge is asymptotically inadmissible. The popular Truncated SVD $\hat{X}_r$ is
asymptotically inadmissible, too:

\begin{thm} \label{inadmis-tsvd:thm}
  {\bf Asymptotic inadmissibility of TSVD.} 
  Let $0<\beta\leq 1$. For any $\rk\in\mathbb{N}$ and any $\V{x}\in\R^\rk$, the
  AMSE of the TSVD estimator $\hat{X}_r$ is well defined, and
  \begin{eqnarray}
    \MSE(\hat{X}_{\lambda_*},\V{x}) \leq 
    \MSE(\hat{X}_\rk,\V{x})\,.
  \end{eqnarray}
   Moreover, strict inequality holds at least at one point $\V{x}_*(\lambda)\in\R^r$.
\end{thm}

\begin{figure*}[!t]
  \centering
    \includegraphics[width=7.7in, trim= 120 0 0 0, clip=true]{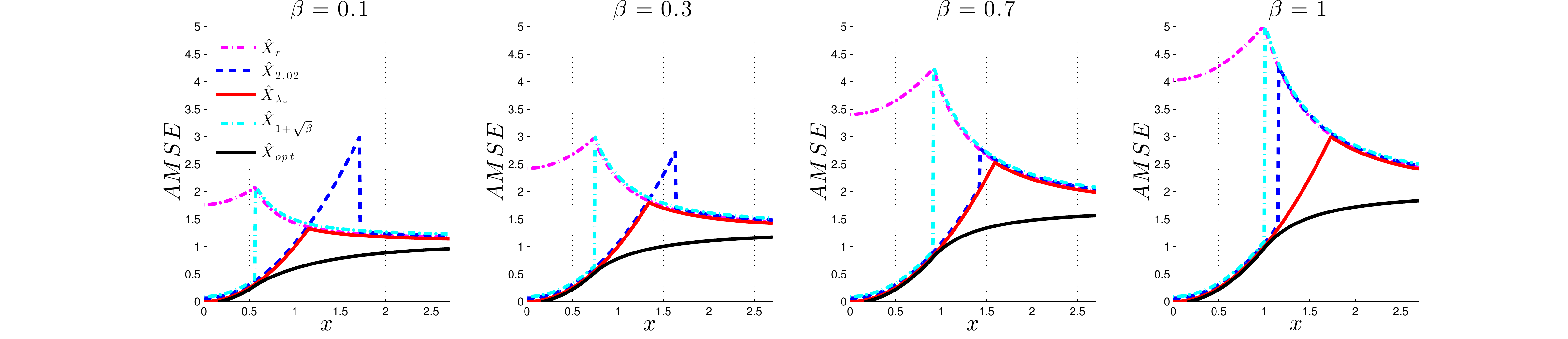}
    \caption{\scriptsize AMSE against signal amplitude $x$ 
  for denoisers discussed: TSVD $\hat{X}_\rk$, universal hard threshold
  $\hat{X}_{2.02}$ from \cite{Chatterjee2010}, 
  and optimally tuned SVHT proposed here $\hat{X}_{\lambda_*}$.
  Also shown: the limiting AMSE of $\hat{X}_\lambda$ as $\lambda\to
  1+\sqrt{\beta}$
  (denoted $\hat{X}_{1+\sqrt{\beta}}$), 
  and optimal singular value shrinkage
  $\hat{X}_{opt}$ from \cite{DonohoGavish2013a}.
  Different aspect ratios $\beta$ are shown;
  $r=1$ everywhere; curves jittered in 
  vertical axis to avoid overlap. }
\label{AMSE:fig}
\end{figure*}

Figure \ref{AMSE:fig} shows the uniform ordering of the 
AMSE curves, stated in Theorems \ref{inadmis:thm} and \ref{inadmis-tsvd:thm},
for a few values of $\beta$. 

To apply the optimal hard threshold to $m$-by-$n$ matrices sampled from the
general model $Y=X+\sigma Z$, by translating $\hat{X}_{\lambda_*}$ using Eq.
\eqref{scale-xhat:eq}, we find the optimal threshold \[\tau_*=\lambda_* \cdot
\sqrt{n}\sigma\,.\] Note that Theorem \ref{inadmis:thm} obviously does not imply
that for any finite matrix $X$ and $\tau\neq \tau_*$ we have
$\norm{\hat{X}_{\tau_*}(X)-X}_F^2 \leq \norm{\hat{X}_{\tau}(X)-X}_F^2$.
However, empirical evidence discussed in Section \ref{MSEvsAMSE:subsec} suggests
that even for relatively small matrices, e.g $n\sim 20$, the performance gain
from using $\hat{X}_{\tau_*}$ is noticeable, and becomes substantial in low SNR.

\subsection{Minimaxity over matrices of bounded rank}

Theorem \ref{inadmis:thm} implies that $\hat{X}_{\lambda_*}$
is asymptotically minimax among SVHT denoisers,
over the class of matrices of a given low rank. Our next result explicitly 
characterizes the least favorable signal
and the asymptotic minimax MSE. 
\begin{thm} \label{mmx-square:thm}
  In the asymptotic square case $\beta=1$, the following holds.
  \begin{enumerate}
    \item {\bf Asymptotically Least Favorable signal for SVHT.}
  Let $\lambda> 2$. Then 
  \begin{eqnarray}
    \text{argmax}_{\V{x}\in\R^\rk} \MSE(\hat{X}_\lambda,\V{x}) =
    x_*(\lambda)\cdot(1,\ldots,1)\in\R^\rk\,,
  \end{eqnarray}
  where 
  \begin{eqnarray*}
    x_*(\lambda) = \frac{\lambda + \sqrt{\lambda^2-4}}{2}\,.    
  \end{eqnarray*}

    \item
  {\bf Minimax AMSE of SVHT.}
For the AMSE of the SVHT denoiser \eqref{hard:eq} we have
    \begin{eqnarray}
      \min_{\lambda > 2} \max_{\V{x}\in\R^\rk} \MSE(\hat{X}_\lambda,\V{x}) &=& 
      3\,\rk \,.
  \end{eqnarray}

\item {\bf Asymptotically minimax tuning of SVHT threshold.}
For the AMSE of the SVHT denoiser \eqref{hard:eq} we have
    \begin{eqnarray}
      \text{argmin}_{\lambda > 2} \max_{\V{x}\in\R^\rk} \MSE(\hat{X}_\lambda,\V{x}) &=& 
      \frac{4}{\sqrt{3}}\,.
  \end{eqnarray}
  \end{enumerate}
\end{thm}

In words, in our asymptotic framework, the least favorable signal for SVHT is
fully degenerate. We will see in Lemma \ref{y-asy:lem} below that the least
favorable location for signal singular values, $x_*(\lambda)$, is such that the
top $r$ observed data singular values fall exactly on the chosen threshold
$\lambda$. 

\begin{thm} \label{mmx-nonsquare:thm}
  For a general asymptotic aspect ratio $0<\beta\leq 1$, the following holds.
  Let $\lambda> 1+\sqrt{\beta}$, then
  \begin{eqnarray}
     \text{argmax}_{\V{x}\in\R^\rk} \MSE(\hat{X}_\lambda,\V{x}) = 
    x_*(\lambda) \cdot(1,\ldots,1)\in\R^\rk \,,
  \end{eqnarray}
  where
  \begin{eqnarray} \label{xstar:eq}
    x_*(\lambda) =
    \sqrt{\frac{\lambda^2-\beta-1+\sqrt{(\lambda^2-\beta-1)^2-4\beta}}{2}}\,.
  \end{eqnarray}
  Moreover,
  \begin{eqnarray} 
    \min_{\lambda>1+\sqrt{\beta}} \,\max_{\V{x}\in\R^\rk} \MSE(\hat{X}_\lambda,\V{x}) =
   \frac{\rk}{2}\cdot\left[(\beta+1) + \sqrt{\beta^2+14\beta+1}\right] 
 \end{eqnarray}
 and
 \begin{IEEEeqnarray}{lCl}
   \IEEEeqnarraymulticol{3}{l}{
     \text{argmin}_{\lambda>1+\sqrt{\beta}}\, \max_{\V{x}\in\R^\rk}
   \MSE(\hat{X}_\lambda,\V{x}) =} 
      \label{mmx-lambda:eq}
   \\\qquad \qquad  &&  
  \sqrt{2(\beta+1)+\frac{8\beta}{(\beta+1)+\sqrt{\beta^2+14 \beta+1}}}
\nonumber\,.
\end{IEEEeqnarray}
\end{thm}

\subsection{Comparison of worst-case AMSE}

By Theorem \ref{inadmis:thm}, the AMSE of optimally tuned SVHT
$\hat{X}_{\lambda_*}$ is always lower than the AMSE of other choices for the
hard threshold location. One way to measure how much worse the other choices
are, and to compare $\hat{X}_{\lambda_*}$ with other popular matrix denoisers,
is to evaluate their worst-case AMSE. 

Table \ref{mmx-square-boundedrank:tab} compares the guarantees provided on AMSE
by shrinkage rules mentioned, for the square matrix case $m=n$ in the model
$Y=X+Z/\sqrt{n}$. For the general noise $Y=X+\sigma Z$ multiply each guarantee
by $n \sigma^2$.

\subsubsection{TSVD}
The AMSE of the TSVD $\hat{X}_\rk$ is calculated in Lemma 
\ref{amse-tsvd:lem} 
below. A simple calculation shows that, in the square matrix case ($\beta=1$)
  \[
  \max_{\V{x}\in\R^\rk} \MSE(\hat{X}_\rk,\V{x})=5\rk\,.
  \]
This is $5/3$ times 
the corresponding worst-case AMSE of $\hat{X}_{\lambda_*}$.

\subsubsection{Hard Thresholding near the bulk edge}

Lemma \ref{amse:lem} provides the AMSE of the SVHT denoiser $\hat{X}_\lambda$,
for any $\lambda>1+\sqrt{\beta}$. A simple calculation shows that 
 \[
 \max_{\V{x}\in\R^\rk} \MSE(\hat{X}_{2.02},\V{x})=4.26\rk\,,
  \]
providing the worst-case AMSE of the Universal Singular Value
Threshold (USVT) of \cite{Chatterjee2010}.
When thresholding near the bulk edge $2$, the change in worse-case AMSE for just a small increase in the threshold
$\lambda$ is drastic (see Figure \ref{AMSE:fig}). The reason for this
phenomenon is discussed in section \ref{discussion:sec}.

\subsubsection{Soft Thresholding} 
Many authors have considered matrix denoising
by applying the soft thresholding nonlinearity $\eta_S(y; s) = (|y|-s)_+ \cdot
\sgn(y)$, instead of hard thresholding, to the data singular values.  The
denoiser  \[ \hat{X}_{s} =\sum_{i=1}^n \eta_{S}(y_i;s) \V{u}_i \V{v}_i' \] is
known as Singular Value Soft Thresholding (SVST) or SVT; See
\cite{Cai2008,Cand2012,DonohoGavish2013} and references therein.  In our
asymptotic framework, following reasoning similar to the proof of Theorem
\ref{inadmis:thm}, one finds that the AMSE of SVST is well defined, and that the
optimal (namely, asymptotically unique admissible) tuning $s_*$ for the soft
threshold is exactly at the bulk edge $1+\sqrt{\beta}$. In the square case, the
AMSE guarantee of optimally-tuned SVST $\hat{X}_{s_*}$ turns out to be $6 \rk$.
This is twice as large as that for the optimally tuned SVHT
$\hat{X}_{\lambda_*}$.  It is interesting to now that both optimal tuning for
the soft threshold $\lambda$ and the corresponding best-possible AMSE guarantee
agree with calculations done in an altogether different asymptotic model, in
which one first takes $n\to\infty$ with rank $r/n\to\rho$, and only then takes
$\rho\to 0$ \cite[sec. 8]{DonohoGavish2013}.  We also note that the worst-case
AMSE of SVST is obtained in the limit of very high SNR, where SVHT does very
well in comparison. When both are optimally tuned, SVHT does not dominate SVST
across all matrices;  In fact, soft thresholding does better than hard
thresholding in low SNR (Figure \ref{convex-envelope:fig}).  For example, in the
square case, when the signal is near $\sqrt{3}$ (the least favorable location
for $\hat{X}_{\lambda_*}$), the AMSE of $\hat{X}_{s_*}$ is
$(7-8/\sqrt{3})r\approx 2.38r$, compared to $3r$, the worse-case AMSE of
$\hat{X}_{\lambda_*}$. 
 
\subsubsection{Optimal Singular Value Shrinker}

Our focus in this paper is denoising by singular value hard thresholding (SVHT), where
$\hat{X}_\lambda$ acts applying a hard thresholding nonlinearity to each of the data singular
values. As mentioned in the introduction, one may ask how SVHT compares to other
singular value shrinkage denoisers, which use a different nonlinearity that may
be more suitable to the problem at hand.
In a special case of our asymptotic framework,
Perry \cite{Perry2009} and Shabalin and Nobel \cite{Shabalin2010} 
have derived an optimal singular value shrinker $\hat{X}_{opt}$.
Proceeding along this line, in  \cite{DonohoGavish2013a} we explore optimal 
shrinkage of singular values under various loss functions and develop a simple expression for
the optimal shrinkers. Calibrated for the model 
$X+Z/\sqrt{n}$,
in the square setting $m=n$, this shrinker takes the form
\[
\hat{X}_{opt} : \sum_{i=1}^n y_i \V{u}_i \V{v}_i'\mapsto
\sum_{i=1}^n \eta_{opt}(y_i) \V{u}_i \V{v}_i'
\]
where
\[
\eta_{opt}(x) = \sqrt{(x^2 - 4)_+}\,.
\]
In our asymptotic framework, this rule dominates in AMSE
essentially {\em any other} estimator based on singular value shrinkage, at any
configuration of the non-zero signal singular values $\V{x}$.
The AMSE of the optimal shrinker (in the square matrix case)
at $\V{x}\in\R^r$ is \cite{DonohoGavish2013a} 
\begin{eqnarray}
  \label{opt:eq}
\MSE(\hat{X}_{opt},\V{x}) = \sum_{i=1}^r 
\begin{cases}
2- \frac{1}{x_i^2} &  x_i \geq 1 \\
x_i^2 & 0\leq x_i \leq 1
\end{cases}\,.
\end{eqnarray}
(See Figure \ref{AMSE:fig}.)
It follows that the worst-case AMSE of $\hat{X}_{opt}$ is 
\[\max_{\V{x}\in\R^\rk} \MSE(\hat{X}_{opt},\V{x})=2r\,\]
 in the square case. We conclude that, for square matrices, in worst-case AMSE,
 singular value hard thresholding at the optimal location is 50\% worse than the
 best possible singular value shrinker, Truncated SVD or SVHT just above the
 bulk-edge
 (which roughly equals the widely used Scree-plot elbow truncation) is 250\% worse, and
 singular value soft thresholding is 300\% worse.

\begin{table*}
  \centering
  \begin{tabular}{|c|c|c|}
    \hline 
    Shrinker & Standing notation & Guarantee on AMSE \\\hline
    Optimal singular value shrinker &$\hat{X}_{opt}$ & $2r$ \\
    Optimally tuned SVHT & $\hat{X}_{\lambda_*}$ & $3r$ \\ 
    Universal Singular Value Threshold \cite{Chatterjee2010} & $\hat{X}_{2.02}$
    & $\approx 4.26r$ \\ 
    TSVD&  $\hat{X}_{r}$ & $5r$ \\ 
    Optimally tuned SVST & $\hat{X}_{s_*}$ & $6r$ \\ \hline 
  \end{tabular}
  \caption{\scriptsize A comparison of guarantees on AMSE provided by singular value
  shrinkage rules discussed, for the square matrix case $m=n$ in the model
$Y=X+Z/\sqrt{n}$. For the general model $Y=X+\sigma Z$ multiply each guarantee
by $n\sigma^2$. }
  \label{mmx-square-boundedrank:tab}
\end{table*}

\subsection{Minimaxity over matrices of bounded nuclear norm}

So far we have considered minimaxity over the class of matrices of at most rank
$r$, where $r$ is given.  In \cite{Chatterjee2010}, the author considered
minimax estimation over a different class of matrices, namely nuclear norm
balls. For a given constant $\xi$, this is the class of all matrices for which
the nuclear norm is at most $\xi$.  Recall that the nuclear norm of a matrix
$X\in\Mmn$, whose vector of singular values is $\V{x}\in\R^m$, is given by
$\norm{\V{x}}_1$.  Our next result shows that $\hat{X}_{\lambda_*}$ is minimax
optimal over this class as well.  Specifically, it is the minimax estimator, in
AMSE, among all SVHT rules, over a given Nuclear Norm ball. We note that unlike
Theorems \ref{mmx-square:thm} and \ref{mmx-nonsquare:thm}, this result does not
follow directly from Theorem \ref{inadmis:thm}. We restrict our discussion to
square matrices ($\beta=1$); the general nonsquare case is handled similarly.

\begin{thm} \label{mmx-l1:thm}
     Let $\lambda>2$ and let $\xi = \rk\cdot (\lambda+\sqrt{\lambda^2-4})/2$ for
     some $\rk\in\mathbb{N}$.     
     \begin{enumerate}
\item The least favorable singular value configuration
     obeys
  \begin{eqnarray}
    \text{argmax}_{\norm{\V{x}}_1\leq \xi} \MSE(\hat{X}_\lambda,\V{x}) =
    x_*(\lambda)\cdot(1,\ldots,1)\in\R^\rk\,,
  \end{eqnarray}
  where 
  \begin{eqnarray*}
    x_*(\lambda) = \frac{\lambda + \sqrt{\lambda^2-4}}{2}\,.    
  \end{eqnarray*}

    \item The best achievable inequality between nuclear norm $\xi$ and AMSE of a hard threshold rule is:
    \begin{eqnarray} \label{eq:BestAcheivable}
      \min_{\lambda>2} \max_{\norm{\V{x}}_1\leq \xi} \MSE(\hat{X}_\lambda,\V{x}) &=& 
      \sqrt{3}\cdot\xi \,.
  \end{eqnarray}
\item The threshold achieving this inequality is 
    \begin{eqnarray}
      \text{argmin}_{\lambda>2} \max_{\norm{\V{x}}_1\leq \xi} \MSE(\hat{X}_\lambda,\V{x}) &=& 
      \frac{4}{\sqrt{3}}\,.
  \end{eqnarray}
  \end{enumerate}
\end{thm}

\begin{figure}[!t]
  \centering
    \includegraphics[width=3.7in]{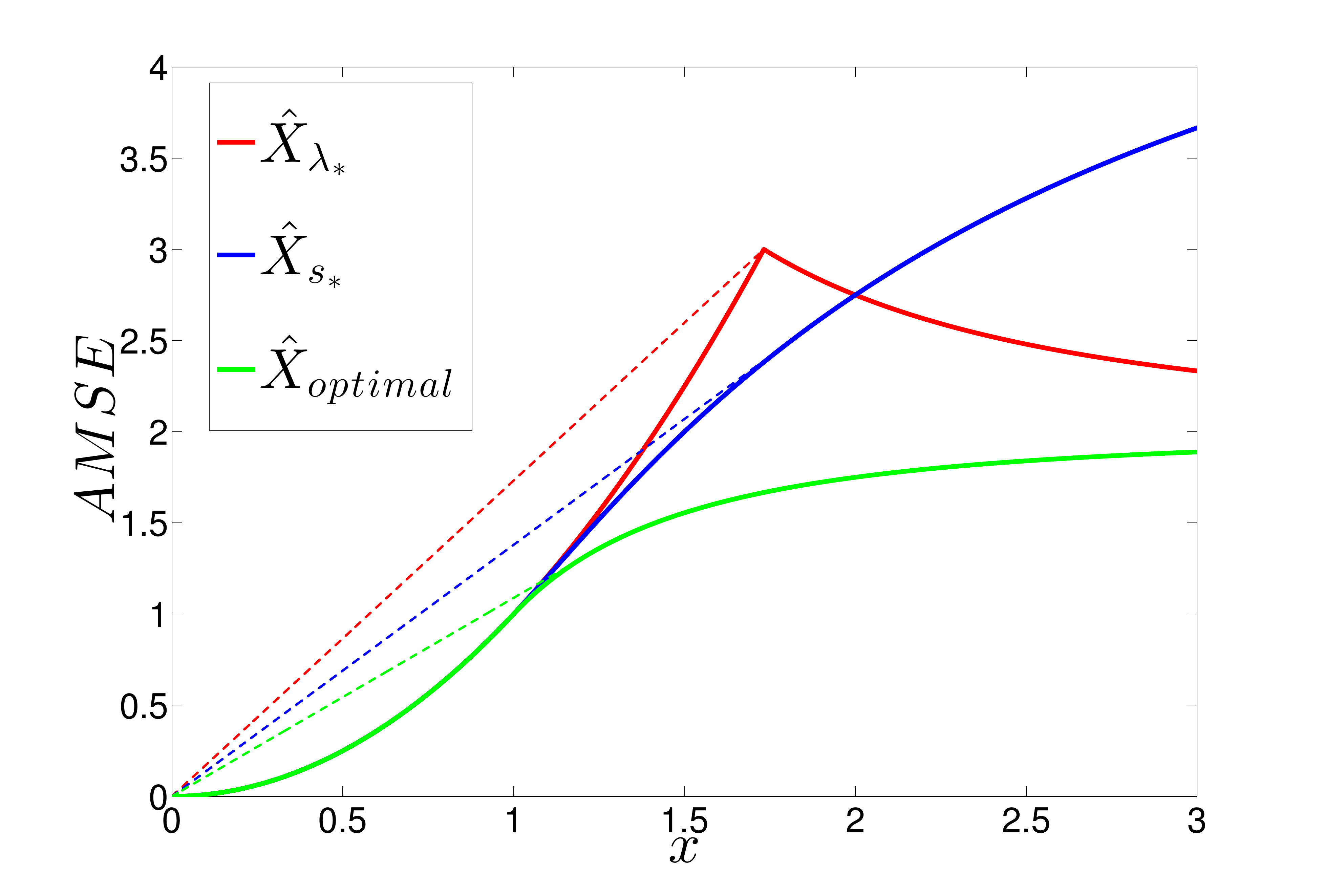}
  \caption{\scriptsize AMSE of optimally tuned SVHT (red), optimally tuned SVST
  (blue) and
  the optimal singular value shrinker \cite{DonohoGavish2013a} (green), 
  for square case ($\beta=1$) and $r=1$.
 The best available constant from Table \ref{mmx-square-l1:tab} is the slope of
 the convex envelope (dashed) of the AMSE curve (solid), namely, the slope of
 the secant running from the origin to the inflection point of the AMSE
 curve. Although the maximum (worst-case AMSE) of optimally tuned SVST $\hat{X}_{s_*}$ is higher
 than that of optimally tuned SVHT $\hat{X}_{\lambda_*}$, the slope of its
 convex envelope is lower. }
  \label{convex-envelope:fig}
\end{figure}

As an alternative to comparing denoisers by comparing their guarantees on AMSE over a prescribed rank $\rk$, one can compare denoisers based on the best available constant $C$ in the
inequality
\begin{eqnarray} \label{eq:BestAcheivable}
  \min_{\lambda>2} \max_{\norm{\V{x}}_1\leq \xi} \MSE(\hat{X}_\lambda,\V{x}) &=& 
      C\cdot\xi \,.
  \end{eqnarray}
The results in the square matrix case are summarized in Table
\ref{mmx-square-l1:tab}. Each constant is derived from the AMSE formula for
 the respective denoiser, as cited above.
To understand why the best available constant for
optimally tuned SVST is smaller than than of optimally tuned SVHT, consider
Figure \ref{convex-envelope:fig}. 

\begin{table*}
\caption{\scriptsize A comparison of best available constant 
  in minimax AMSE, over nuclear norm balls, for the shrinkage rules discussed, 
  for the square matrix case $m=n$ in the model
$Y=X+Z/\sqrt{n}$. These constants are the same for the general model $Y=X+\sigma
Z$.}
  \centering
  \begin{tabular}{|c|c|c|}
    \hline 
    Shrinker & Standing notation & Best possible constant $C$ in Eq.
    \eqref{eq:BestAcheivable} \\\hline
    Optimal singular value shrinker &$\hat{X}_{opt}$ & $1$ \\
    Optimally tuned SVHT & $\hat{X}_{\lambda_*}$ & $\sqrt{3}\approx 1.73$ \\ 
    USVT of \cite{Chatterjee2010} & $\hat{X}_{2.02}$ & $\approx 3.70$ \\ 
    TSVD&  $\hat{X}_{r}$ & $5$ \\ 
    Optimally tuned SVST & $\hat{X}_{s_*}$ &  $ \approx 1.38$ \\ \hline 
  \end{tabular}
    \label{mmx-square-l1:tab}
\end{table*}

\subsection{When the noise level $\sigma$ is unknown} \label{sigmahat:subsec}

When the noise level in which $Y$ is observed is unknown, it no longer makes sense to use
$\hat{X}_{\lambda_*}$, which is calibrated for a specific noise level. 
We now
describe a method to estimate the optimal hard threshold from the data matrix
$Y$. To emphasize that the resulting denoiser is ready for use on data from the
general model $Y=X+\sigma Z$, we denote this estimated threshold by
$\hat{\tau}_*$, and the SVHT denoiser by $\hat{X}_{\hat{\tau}_*}$.
To this end, we are required to estimate the unknown noise level $\sigma$.
In the closely related Spiked Covariance Model, there are existing methods for
estimation of an unknown noise level; see for example \cite{Kritchman2009} and
references therein.

Consider the following robust estimator for the
parameter $\sigma$ in the model $Y=X+\sigma Z$:
\begin{eqnarray}
  \hat{\sigma}(Y) \eqdef \frac{y_{med}}{\sqrt{n \cdot\mu_\beta}}\,,
\end{eqnarray}
where $y_{med}$ is a median singular value of $Y$ and $\mu_\beta$ is the
median of the the Mar\v{c}enko-Pastur distribution, namely, the unique solution 
in $\beta_-\leq x\leq \beta_+$
to the equation
\begin{eqnarray*}
  \intop_{\beta_-}^x \frac{\sqrt{(\beta_+-t)(t-\beta_-)}}{2\pi t}dt =
  \frac{1}{2}\,,
\end{eqnarray*}
where $\beta_\pm=(1\pm\sqrt{\beta})^2$.
Define the optimal hard threshold for a data matrix $Y\in\Mmn$ observed in unknown noise
level, with $m/n=\beta$,
by plugging in $\hat{\sigma}(Y)$ instead of $\sigma$ in Eq.
\eqref{taustar-nonsq:eq}:
\[
\hat{\tau}_*(\beta,Y) \eqdef \lambda_*(\beta) \cdot \sqrt{n}\, \hat{\sigma}(Y) =
\frac{\lambda_*(\beta)}{\sqrt{\mu_\beta}} y_{med}\,.
\]
Writing $\omega(\beta) = \lambda_*(\beta)/\sqrt{\mu_\beta}$, the threshold is 
\[
\hat{\tau}_*(\beta,Y) = \omega(\beta) \cdot y_{med}\,.
\]

The median $\mu_\beta$ and hence the coefficient $\omega(\beta)$ are not available analytically; 
in \cite{code} we make available a 
Matlab script to 
evaluate the coefficient  $\omega(\beta)$.
Some values are tabulated in 
Table \ref{lambdastar-hat:tab} for convenience.
A useful approximation to $\omega$ is given as a cubic polynomial in  Eq.
\eqref{omega-approx:eq} above.
Empirically, 
\begin{eqnarray*}
\max_{0.001<\beta\leq 1 }|\omega(\beta) -
   \left(0.56\beta^3 - 0.95\beta^2 + 1.82\beta + 1.43\right)| 
\leq 0.02
\end{eqnarray*}
which may be sufficient for some practical purposes if one does not have access to
a more exact value of $\omega$.

\begin{lemma} \label{sigmahat:lem}
  For the sequence $Y_n$ in our asymptotic framework, 
  \[\lim_{n\to\infty} \frac{\hat{\sigma}(Y_n)}{1/\sqrt{n}}\aseq  1 .\]
\end{lemma}
\begin{cor}
For $Y_n$ as above and any $0<\beta\leq 1$, 
\[ 
\lim_{n\to\infty} \hat{\tau}_*(\beta,Y_n) \aseq 
 \lambda_*(\beta)\cdot\lim_{n\to\infty} \sqrt{n}\cdot \hat{\sigma}(Y_n) \aseq
\lambda_*(\beta)\,,
\]
\end{cor}
\begin{cor}
For $Y_n$ as above, any $0<\beta\leq 1$,
any $r$ and any $\V{x}\in\R^r$, almost surely
\[
\MSE(\hat{X}_{\hat{\tau}_*},\V{x}) =
\MSE(\hat{X}_{\lambda_*},\V{x})\,.
\]
\end{cor}
\begin{cor}
 Theorem \ref{inadmis:thm}, Theorem \ref{inadmis-tsvd:thm},
Theorem \ref{mmx-square:thm}, Theorem \ref{mmx-nonsquare:thm} and Theorem
\ref{mmx-l1:thm} all hold if we replace the optimally-tuned SVHT for known
$\sigma$,
$\hat{X}_{\lambda_*(\beta)}$, by optimally-tuned SVHT for unknown $\sigma$,
$\hat{X}_{\hat{\tau}_*}$.
\end{cor}

\begin{table}
  \centering
    \begin{tabular}{||r|l||r|l||}
\hline
$\beta$ & $\omega(\beta)$& $\beta$ &
$\omega(\beta)$ \\\hline\hline
0.05&1.5194&0.55&2.2365\\\hline
0.10&1.6089&0.60&2.3021\\\hline
0.15&1.6896&0.65&2.3679\\\hline
0.20&1.7650&0.70&2.4339\\\hline
0.25&1.8371&0.75&2.5011\\\hline
0.30&1.9061&0.80&2.5697\\\hline
0.35&1.9741&0.85&2.6399\\\hline
0.40&2.0403&0.90&2.7099\\\hline
0.45&2.106&0.95&2.7832\\\hline
0.50&2.1711&1.00&2.8582\\\hline
\end{tabular}
\caption{\scriptsize Some values of the optimal hard threshold coefficient 
for unknown noise level, $\omega(\beta)$ of Eq. \eqref{tauhatstar-nonsq:eq}
   For an
  $m$-by-$n$ matrix in unknown noise level (with $m/n=\beta$), the 
  optimal SVHT denoiser $\hat{X}_{\hat{\tau}_*}$ sets to zero 
  all data singular values below the threshold
  $\tau_*=\omega(\beta) y_{med}$, where $y_{med}$ is the median singular
  value of the data matrix $Y$. Calculated using function provided in the code
  supplement
  \cite{code}.}
  \label{lambdastar-hat:tab}
\end{table}

\section{Discussion}
\label{discussion:sec}

\subsection{The optimal threshold $\lambda_*(\beta)$ and the bulk edge
$1+\sqrt{\beta}$}

\begin{figure}[!t]
  \centering
    \includegraphics[width=3.4in]{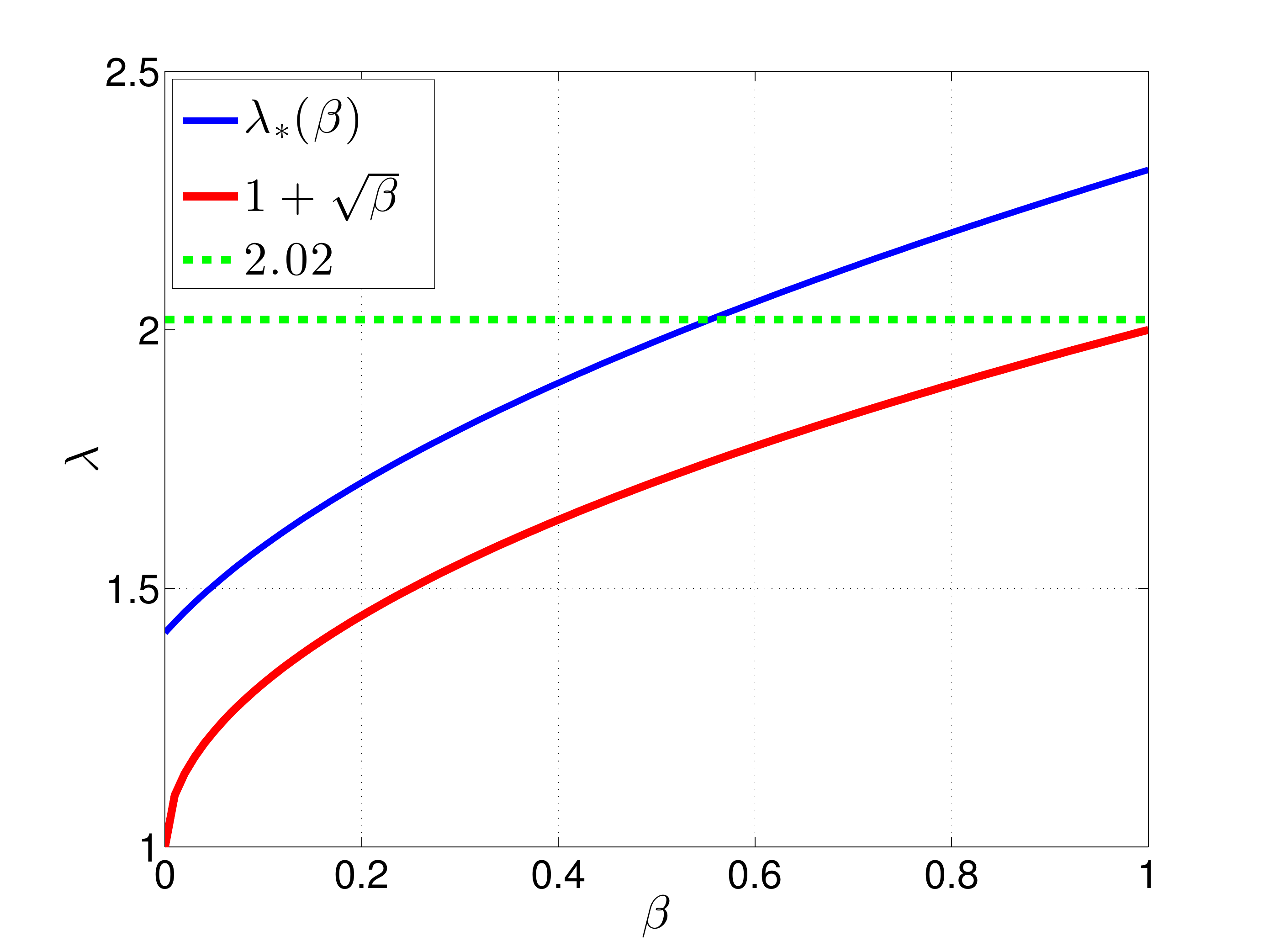}
    \caption{\scriptsize The optimal threshold $\lambda_*(\beta)$ from
  \eqref{lambdastar-nonsquare:eq}
  against $\beta$. Also shown are bulk-edge
  $1+\sqrt{\beta}$, which is the hard threshold corresponding to TSVD in our
  setting, and the USVT threshold $2.02$ from  \cite{Chatterjee2010}. }
\label{lambda:fig}
\end{figure}

Figure \ref{lambda:fig} shows the optimal threshold $\lambda_*(\beta)$ over
$\beta$.
The edge of the quarter circle bulk $1+\sqrt{\beta}$,  the hard threshold that 
best emulates TSVD in our setting,  is shown for
comparison. 
In the null case $X=0$, the largest data singular value is located
asymptotically exactly
at the bulk edge, $1 + \sqrt{\beta}$.  It might seem that just above 
the bulk edge is
a natural place to set
a threshold, since anything smaller could be the product of a pure noise situation. 
However, 
for $\beta > 0.2$, the optimal hard threshold $\lambda^*(\beta)$ is 15-20\% larger than the bulk edge;
as $\beta \goto 0$, it grows about 40\% larger.
Inspecting the proof of Theorem \ref{inadmis:thm} and particularly the
expression for AMSE of SVHT (Lemma \ref{amse:lem}), one finds the reason: 
one component of the AMSE is due to the angle between the signal singular
vectors and the data singular vectors. This angle converges to a nonzero value
as $n\to\infty$ (given explicitly in Lemma \ref{inner-asy:lem}) which grows as SNR decreases. 
When some data singular value $y_i$ is too
close to the bulk, its corresponding singular vectors are too badly rotated, 
and the rank-one matrix $y_i \V{u}_i \V{v}_i'$ it contributes to
the denoiser hurts the AMSE more than it helps. For example, for square matrices
$\beta=1$, this situation is most
acute when the signal singular value is just barely larger than $x_i=1$, causing the corresponding data
singular value $y_i$ to be just barely larger than the bulk edge, which for
square matrices is located at $2$. A SVHT denoiser thresholding just above 
the bulk edge
would include the component $y_i \V{u}_i\V{v}_i'$, incurring an AMSE 
about 5 times larger than the AMSE incurred by
excluding $y_i$ from the reconstruction.
The optimal threshold
$\lambda_*(\beta)$ keeps such singular values out of the picture; this is why it
is necessarily larger than the bulk edge. The precise value of
$\lambda_*(\beta)$ is the precise point at which it becomes advantageous to
include the rank-one contribution of a singular value $y_i$ in the 
reconstruction.  

\subsection{The optimal threshold $\lambda_*(\beta)$ relative to the USVT $\hat{X}_{2.02}$}

As mentioned in the introduction,  S. Chatterjee has recently discussed SVHT in a broad class of
situations \cite{Chatterjee2010}. Translating his much broader discussion to the
confines of the present context, he observed that {\em any} 
$\lambda>2$  can serve as a {\em universal hard threshold for singular values}
(USVT), offering fairly good
performance regardless of the matrix shape $\m/\n$ and the underlying signal
matrix $X$. The author makes the specific recommendation $\lambda=2.02$ and writes:
\begin{quote} {\em ``The algorithm manages to cut off the singular values at the `correct'
level, depending on the structure of the unknown parameter matrix. 
The adaptiveness of the USVT threshold is somewhat similar in spirit to that of
the SureShrink algorithm of Donoho and Johnstone.``}
\end{quote}

Keeping in mind that the scope of \cite{Chatterjee2010} is much broader than the
one considered here, we would like to evaluate this proposal, in the setting of
low rank matrix in white noise, and specifically in our
asymptotic framework.
Figure \ref{lambda:fig} includes the value $2.02$: indeed, this threshold is
larger than the bulk edge, for any $0<\beta\leq 1$, so
Chatterjee's $\hat{X}_{2.02}$ rule asymptotically set to zero all
singular values which {\em could} arise 
due to an underlying noise-only situation. When $\lambda_*(\beta)<2.02$, the
$\hat{X}_{2.02}$ rule sometimes ``kills'' singular values that the optimal threshold
deems good enough for keeping, and when $\lambda_*(\beta)>2.02$, the
$\hat{X}_{2.02}$ rule sometimes ``keeps'' singular values that did in fact arise from
signal, but are so close to the bulk that the optimal threshold declares them
unusable.

For $\beta=1$, the guarantee on worst-case AMSE obtained by using $\lambda=2.02$
over matrices of rank $\rk$ is about $4.26\rk$, roughly 
140\% larger than the guarantee obtained by using the minimax threshold 
$\lambda=4/\sqrt{3}$ (See Figure \ref{AMSE:fig}). For square matrices, the
regret for preferring USVT to optimally-tuned SVHT can be substantial: in low SNR 
($x\approx 1$), using the threshold $\lambda=2.02$ incurs roughly
twice the 
AMSE  of the minimax threshold $4/\sqrt{3}$. 

\begin{figure}[!t]
  \centering
    \includegraphics[width=3.4in]{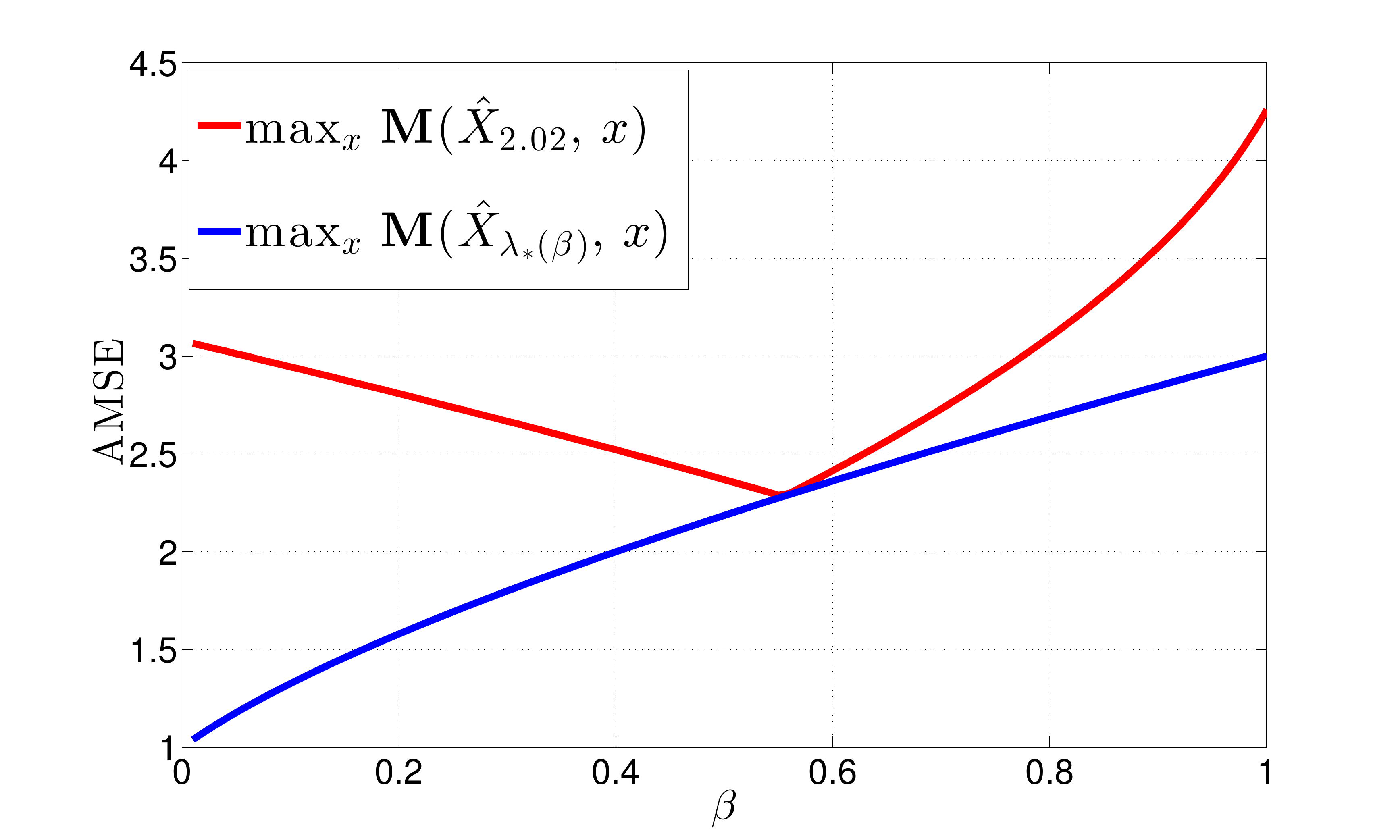}
    \caption{\scriptsize Worst-case AMSE against the shape parameter 
  $\beta$, 
  for two choices of hard threshold:
  $\lambda=2.02$ from \cite{Chatterjee2010},
  and optimal threshold $\lambda_*$  from
  \eqref{lambdastar-nonsquare:eq}. }
\label{sourav-mmx-lf:fig}

\end{figure}

We note that unlike
 the optimally tuned SVHT
$\hat{X}_{\lambda_*}$, the USVT
$\hat{X}_{2.02}$ does not take into account the shape factor $\beta$, namely the
ratio of number of rows to number of columns of the matrix in question.
A comparison of worst-case AMSE between the fixed threshold choice
$\lambda=2.02$ and
the optimal hard threshold
$\lambda=\lambda_*(\beta)$ is shown in Figure \ref{sourav-mmx-lf:fig}.
The two curves intersect at $\lambda\approx 0.55$,
where the optimal threshold \eqref{lambdastar-nonsquare:eq}
is approximately $2.02$.

One might argue that  \cite{Chatterjee2010} proposed $2.02$ based on
its MSE performance over classes of matrices bounded in nuclear norm.
But also for that purpose, $2.02$ is noticeably outperformed by $\lambda^*(\beta)$.
Arguing as in Theorem \ref{mmx-l1:thm} we obtain, in the square case:
    \begin{eqnarray}
\max_{\norm{\V{x}}_1\leq \xi} \MSE(\hat{X}_{2.02},\V{x}) &\approx& 
      3.70 \cdot\xi \,.
\end{eqnarray}
The coefficient $3.70$ is about 110\% larger than 
the best coefficient achievable by SVHT, namely  $C=\sqrt{3}$ in \eqref{eq:BestAcheivable}.

One should keep in mind that
USVT is applicable for a wide range of noise models, e.g. in stochastic block
models.
\cite{Chatterjee2010} is the first, to the best of out knowledge, to suggest
that a matrix denoising procedure as simple as SVHT could have universal
optimality properties. In our asymptotic framework of low-rank matrices in white
noise,  the 2.02 threshold performs fairly well in AMSE, except for very small
values of $\beta$ (Figure \ref{AMSE:fig});
but one often gets a substantial  AMSE improvement by switching to the
rule we recommend. Since our recommendation dominates in AMSE,
there is no downside to making this switch -- i.e. there is no configuration
of signal singular values $x$ which could make one regret this switch.

\section{Proofs}

Setting additional notation required in the proofs, let 

\[X_n=\sum_{i=1}^r x_i \, \V{a}_{n,i} \, \V{b}_{n,i}'
\] be a sequence of signal
matrices in our asymptotic framework, so that $\V{a}_{n,i}\in\R^{m_n}$ (resp.
$\V{b}_{n,i}\in\R^n$) is the
left (resp. right) singular vector corresponding to the singular value $x_i$,
namely, 
$i$-th column of $U_n$ (resp. $V_n$) in \eqref{singvec:eq}.
Similarly, let $Y_\n$ be a corresponding sequence of observed matrices in our
framework, and write 
\[
  Y_n = \sum_{i=1}^{m_n} y_{n,i} \,\V{u}_{n,i}\, \V{v}_{n,i}'
\]
so that $\V{u}_{n,i}\in\R^m$ (resp. $\V{v}_{n,i}\in\R^n$) is the
left (resp. right) singular vector corresponding to the singular value $y_{n,i}$.
(Note that $\left\{ \V{a}_{n,i} \right\}$ and $\left\{ \V{b}_{n,i} \right\}$ are
unknown, arbitrary, non-random vectors.)

Our main results depend on Lemma \ref{amse:lem}, a formula for the AMSE of SVHT.
This formula in turn depends on Lemma \ref{y-asy:lem} and Lemma
\ref{inner-asy:lem}. Both follow from recent key results due to
\cite{Benaych-Georges2012}.

\begin{lemma} \label{y-asy:lem}
  {\bf Asymptotic  data singular values.}
  For $1\leq i\leq \rk$,
  \begin{eqnarray} \label{y:eq}
    \lim_{\n\to\infty} y_{\n,i} \aseq
    \begin{cases}
    \sqrt{\left(x_i +
    \frac{1}{x_i}\right)\left(x_i + \frac{\beta}{x_i}\right)} 
    & x_i>\beta^{1/4} \\
    1+\sqrt{\beta} & x_i\leq\beta^{1/4}
  \end{cases}
  \end{eqnarray}
\end{lemma}

\begin{lemma} \label{inner-asy:lem}
  {\bf Asymptotic angle between signal and data singular vectors.}
  Let $1\leq i\neq j\leq \rk$ and assume that $x_i$ has degeneracy $d$, namely,
  there are exactly $d$ entries of $\V{x}$ equal to $x_i$. If 
$x_i>\beta^{1/4}$, we have
\begin{eqnarray} \label{proj-u:eq}
  d\cdot \lim_{\n\to\infty} \big|\langle \V{a}_{n,i}\,,\,\V{u}_{\n,j}\rangle\big|^2 \stackrel{a.s.}{=} 
    \begin{cases}
\frac{x_i^4 - \beta}{x_i^4 + \beta x_i^2} & x_i = x_j \\
      0 & x_i \neq x_j
    \end{cases}\,,
  \end{eqnarray}
and, a slightly different formula, 
\begin{eqnarray} \label{proj-v:eq}
  d\cdot \lim_{\n\to\infty} \big|\langle \V{b}_{n,i}\,,\,\V{v}_{\n,j}\rangle\big|^2 \stackrel{a.s.}{=} 
    \begin{cases}
      \frac{x_i^4 - \beta}{x_i^4 +  x_i^2} & x_i = x_j \\
      0 & x_i \neq x_j
    \end{cases}\,.
  \end{eqnarray}
  If however $x_i\leq\beta^{1/4}$, then we have
  \begin{eqnarray*}
    \lim_{\n\to\infty} \big|\langle \V{a}_{n,i}\,,\,\V{u}_{\n,j}\rangle\big| \stackrel{a.s.}{=} 
    \lim_{\n\to\infty} \big|\langle \V{b}_{n,i}\,,\,\V{v}_{\n,j}\rangle\big| \stackrel{a.s.}{=} 
0\,.
  \end{eqnarray*}
\end{lemma}

To appeal to these results, we need to show that our asymptotic framework
satisfies the assumptions of  \cite{Benaych-Georges2012}. 
By \cite{silverstein_book} the limiting law of
 the singular values of
$Z_n/\sqrt{n}$ is the quarter-circle density 
\begin{eqnarray} \label{mp:eq}
  f(x) = \frac{\sqrt{4\beta-(x^2-1-\beta)^2}}{\pi \beta
  x}\mathbf{1}_{[1-\sqrt{\beta},1+\sqrt{\beta]}}(x)\,;
\end{eqnarray}
by \cite{Yin1988}, $y_{n,1}\aslim 1+\sqrt{\beta}$; by \cite{Bai1993},
$y_{n,m_n}\aslim 1-\sqrt{\beta}$. This satisfies 
 assumptions
2.1, 2.2 and 2.3 of \cite{Benaych-Georges2012}, respectively.
Formulas \eqref{y:eq}, \eqref{proj-u:eq} and \eqref{proj-v:eq},  as seen in 
\cite[example 3.1]{Benaych-Georges2012},
depend only on the
shape of the limiting distribution \eqref{mp:eq} and not on any Gaussian
assumptions.

Using Lemma \ref{y-asy:lem} and Lemma \ref{inner-asy:lem}, we can calculate the
AMSE \eqref{amse_def:eq} of the hard thresholding estimator $\hat{X}_\lambda$,
for given threshold $\lambda$, at a matrix of specific aspect ratio $\beta$ and
signal singular values $\V{x}$:

\begin{lemma} \label{amse:lem}
  {\bf AMSE of singular value hard thresholding.}
  Fix $\rk>0$ and $\V{x}\in\R^\rk$.
  Let $\{X_\n(\V{x})\}_{\n=1}^\infty$ and
  $\{Z_\n\}_{\n=1}^\infty$ be
  matrix sequences in our asymptotic framework, and let $\lambda\geq 1+\sqrt{\beta}$. 
  Then
  \begin{eqnarray} \label{amse:eq}
    \MSE(\hat{X}_\lambda,\V{x}) = \sum_{i=1}^\rk M(\hat{X}_\lambda,x_i) 
  \end{eqnarray}
  where
  \begin{IEEEeqnarray}{lCl} 
    \IEEEeqnarraymulticol{3}{l}{
      M(\hat{X}_\lambda,x)=} 
      \label{M:eq}
      \\ \qquad &&
    \begin{cases}
      (x+\frac{1}{x})(x + \frac{\beta}{x})- (x^2-\frac{2\beta}{x^2})
      &
 \,\, x \geq x_*(\lambda)\\
 x^2 & 
 \,\, x < x_*(\lambda)
    \end{cases}\,\nonumber
  \end{IEEEeqnarray}
  and $x_*(\lambda)$ is given by Eq. \eqref{xstar:eq}.
\end{lemma}

Figure \ref{AMSE:fig} shows the AMSE of Lemma
\ref{amse:lem}, in square case $\beta=1$ and nonsquare cases $\beta=0.1$, 
$\beta=0.3$ and $\beta=0.7$.

\begin{proof}
  By definition,
  \begin{eqnarray*}
    \hat{X}_\lambda(Y_\n) = \sum_{i=1}^{m_n}
    \eta_H(y_{n,i};\lambda)\,\V{u}_{\n,i}\,\V{v}_{\n,i}' \,,
  \end{eqnarray*}
where $\eta_H(y,\tau) = y \,\mathbf{1}_{\{y \geq \tau \}}$.
  Observe that
  \begin{IEEEeqnarray}{llCl}
    \IEEEeqnarraymulticol{4}{l}{
      \norm{\hat{X}_\lambda(Y_\n)-X_\n}_F^2=} 
    \label{finite-dif:eq} \nonumber\\
    \qquad &&& 
    \inner{\hat{X}_\lambda(Y_\n)-X_n}{\hat{X}_\lambda(Y_\n)-X_n}  =\nonumber  \\
&&&     \inner{\hat{X}_\lambda(Y_n)}{\hat{X}_\lambda(Y_n)}+ \nonumber \\
&&& \qquad\inner{X_\n}{X_\n} -2\inner{\hat{X}_\lambda(Y_\n)}{X_\n}= \nonumber \\
&&& 
\sum_{i=1}^{m_n} \eta_H(y_{\n,i};\lambda)^2 +\sum_{i=1}^\rk x_i^2 -\nonumber\\ 
&&&\qquad 2\sum_{i,j=1}^\rk x_i \, \eta_H(y_{\n,j};\lambda) \,
\inner{\V{a}_i\V{b}_i'}{\V{u}_{\n,j}\V{v}_{\n,j}'} =\nonumber\\
&&& 
\sum_{i=\rk+1}^{m_n} \eta_H(y_{n,i};\lambda)^2 + \sum_{i=1}^\rk\Bigg[ \eta_H(y_{\n,i};\lambda)^2 + x_i^2 -\nonumber \\
&&&\qquad 2 x_i \sum_{j=1}^\rk \eta_H(y_{\n,j};\lambda) \,
\inner{\V{a}_i\V{b}_i'}{\V{u}_{\n,j}\V{v}_{\n,j}'}\Bigg]\,.
  \end{IEEEeqnarray}
  Since  $y_{n,r+1}\aslim 1+\sqrt{\beta} < \lambda$, the leftmost term above converges almost surely to zero. 
  When $0\leq x_i\leq \beta^{1/4}$, by Lemma \ref{y-asy:lem} and Lemma
  \ref{inner-asy:lem},
  only the term $x_i$ survives and Eq.
  \eqref{M:eq} holds. Assume now that $x_i>\beta^{1/4}$.
  We now consider the a.s. limiting value of each of the remaining terms in 
  \eqref{finite-dif:eq}. For the term $\sum_{i=1}^r \eta_H(y_{n,i};\lambda)^2$, by Lemma \ref{y-asy:lem},
  for $i=1,\ldots,\rk$ we have
  \begin{IEEEeqnarray*}{lCl}
    \IEEEeqnarraymulticol{3}{l}{
      \lim_{\n\to\infty} \eta_H(y_{n,i};\lambda)^2 \aseq}
      \\ \qquad && 
    \begin{cases}
 (x_i+\frac{1}{x_i})(x_i + \frac{\beta}{x_i}) &
  (x_i+\frac{1}{x_i})(x_i + \frac{\beta}{x_i}) \geq \lambda^2 \\
  0 &
 (x_i+\frac{1}{x_i})(x_i + \frac{\beta}{x_i}) < \lambda^2
 \end{cases}\,.
  \end{IEEEeqnarray*}

  Turning to the rightmost term of \eqref{finite-dif:eq},  
  by Lemma \ref{inner-asy:lem}, for $i,j=1,\ldots,\rk$ we find that it equals
    \begin{IEEEeqnarray}{lCl}
          \IEEEeqnarraymulticol{3}{l}{
    \lim_{\n\to\infty} 
    \inner{\V{a}_i}{\V{u}_{\n,j}} \inner{\V{b}_i}{\V{v}_{\n,j}'}\aseq} 
    \\ \qquad\qquad
    && 
     \begin{cases}
      \frac{1}{d_i}\frac{x_i^4-\beta}{x_i^3\sqrt{(x_i+\beta/x_i)(x_i+1/x_i)}} &
      x_i = x_j \\
      0 & x_i\neq x_j
    \end{cases} \nonumber
\end{IEEEeqnarray}
  where $d_i = \#\left\{ j\,|\, x_j=x_i \right\}$.
  Furthermore, since for all
  $\V{x}_1,\V{x}_2\in\R^\m$ and $\V{y}_1,\V{y}_2\in\R^\n$ we have
  $\inner{\V{x}_1\V{y}_1'}{\V{x}_2\V{y}_2'}=\inner{\V{x}_1}{\V{x}_2}\inner{\V{y}_1}{\V{y}_2}$, 
  we find that for $i=1,\ldots,\rk$,
  \begin{IEEEeqnarray*}{lCl}
    \IEEEeqnarraymulticol{3}{l}{
    \sum_{j=1}^\rk \eta(y_{\n,j};\lambda) \,
    \inner{\V{a}_i\V{b}_i'}{\V{u}_{\n,j}\V{v}_{\n,j}'} =} 
    \\ \qquad &&  
    \sum_{j=1}^\rk \eta(y_{\n,j};\lambda) \,
    \inner{\V{a}_i}{\V{u}_{\n,j}}
\inner{\V{b}_i}{\V{v}_{\n,j}'}\,. 
  \end{IEEEeqnarray*}
 For the rightmost term of
  \eqref{finite-dif:eq} we conclude that
  \begin{IEEEeqnarray*}{lCl}
    \IEEEeqnarraymulticol{3}{l}{  \lim_{\n\to\infty}
x_i \sum_{j=1}^\rk \eta(y_{\n,j};\lambda) \,
\inner{\V{a}_i\V{}_i'}{\V{u}_{\n,j}\V{v}_{\n,j}'} \aseq}\\
\qquad
&& 
\sum_{1\leq j\leq \rk\,:\,x_j=x_i} \lim_{\n\to\infty}  \cdot 
 \frac{ \eta(y_{\n,j};\lambda) \cdot 
(x_i^4-\beta)}{d_i x_i^2\sqrt{(x_i+\beta/x_i)(x_i+1/x_i)}} =\\
\qquad &&
    \begin{cases}
\frac{x_i^4-\beta}{x_i^2} &
  \,(x_i+\frac{1}{x_i})(x_i + \frac{\beta}{x_i}) \geq \lambda^2 \\
  0 &
 \,(x_i+\frac{1}{x_i})(x_i + \frac{\beta}{x_i}) < \lambda^2
 \end{cases}\,,
  \end{IEEEeqnarray*}
  where we have used Lemma \ref{y-asy:lem} again. 
  Collecting the terms, we find for the limiting value of \eqref{finite-dif:eq} that
  \begin{eqnarray}
    \lim_{\n\to\infty} \norm{\hat{X}_\lambda(Y_\n)- X_\n}_F^2 \aseq
    \sum_{i=1}^\rk M(\hat{X}_\lambda,x_i)\,,
  \end{eqnarray}
  where $M(\hat{X}_\lambda,x)$ is given by \eqref{M:eq} as required. 
  \end{proof}

For the TSVD, the same argument gives:
\begin{lemma} \label{amse-tsvd:lem}
  {\bf AMSE of TSVD.}
  Fix $\rk>0$ and $\V{x}\in\R^\rk$.
  Let $\{X_\n(\V{x})\}_{\n=1}^\infty$ and
  $\{Z_\n\}_{\n=1}^\infty$ be
  matrix sequences in our asymptotic framework, and let $\lambda\geq 1+\sqrt{\beta}$. 
  Then
  \begin{eqnarray} \label{amse-tsvd:eq}
    \MSE(\hat{X}_r,\V{x}) = \sum_{i=1}^\rk M(\hat{X}_r,x_i) 
  \end{eqnarray}
  where
  \begin{IEEEeqnarray}{lCl} 
    \IEEEeqnarraymulticol{3}{l}{
      M(\hat{X}_r,x)=} 
      \label{M_TSVD:eq}
      \\ \qquad && 
    \begin{cases}
      (x+\frac{1}{x})(x + \frac{\beta}{x})- (x^2-\frac{2\beta}{x^2})
      &
      \,\, x \geq \beta^{1/4} \\
(1+\sqrt{\beta})^2 + x^2 & 
\,\, x \leq \beta^{1/4}
    \end{cases}\,.
    \nonumber
  \end{IEEEeqnarray}
\end{lemma}

We now to turn to prove our main results. 

\subsubsection*{Proof of Theorem \ref{inadmis:thm}}
    Let $x_*=x_*(\lambda_*(\beta))$ where $\lambda_*(\beta)$ is defined in 
    \eqref{lambdastar-nonsquare:eq} and $x_*(\lambda)$ is defined in
    \eqref{xstar:eq}. Then
 \[ x_*^2 = 
    \left(x_*+\frac{1}{x_*}\right)\left(x_*+\frac{\beta}{x_*}\right) -
    \left(x_*^2-\frac{2\beta}{x_*^2}\right)
    \,.\]
    It follows that for all $x>0$ and $\lambda\geq 1+\sqrt{\beta}$,
    \begin{IEEEeqnarray*}{lCl}
      \IEEEeqnarraymulticol{3}{l}{
        M(\hat{X}_{\lambda_*},x) } 
        \\ \qquad &=&  \min\left\{ x^2 \,,\, 
    \left(x+\frac{1}{x}\right)\left(x+\frac{\beta}{x}\right) -
    \left(x^2-\frac{2\beta}{x^2}\right) 
    \right\}
    \\
  \qquad &\leq& M(\hat{X}_\lambda,x)
  \end{IEEEeqnarray*}
    and the theorem follows from Eq. \eqref{amse:eq}.
\qed

\begin{figure*}[t!]
  \centering
    \includegraphics[height=1.39in,  trim= 180 1 3 .5, clip=true]{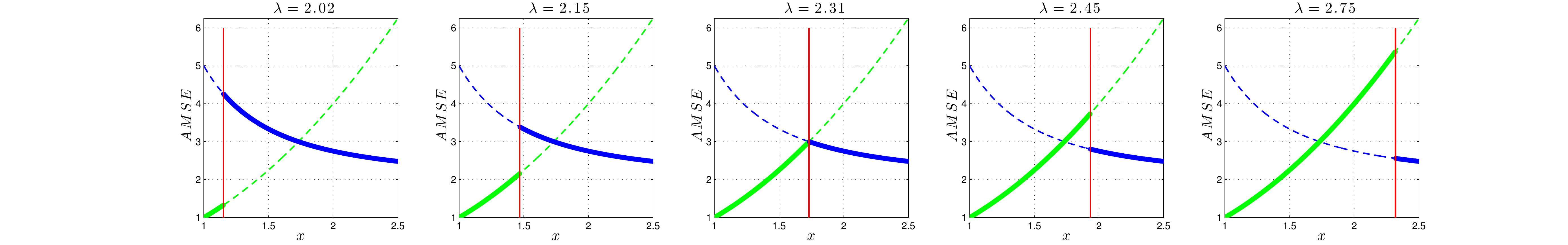}
  \caption{\scriptsize The AMSE profiles of \eqref{amse:eq} for $\beta=1$,
  $\rk=1$ and several threshold values $\lambda$.
  Green: $x\mapsto x^2$. Blue: $x\mapsto 2+3/x^2$. Horizontal line:
  location of the cutoff $x_*(\lambda)$ solving $x_* + 1/x_* = \lambda$. Solid line: 
  AMSE curve. }
  \label{lf:fig}
\end{figure*}

Figure \ref{lf:fig} provides a visual explanation of this proof for the square 
($\beta=1$) case.

\subsubsection*{Proof of Theorem \ref{inadmis-tsvd:thm}}
For $x<\beta^{1/4}$, by Lemma \ref{amse:lem} and Lemma \ref{amse-tsvd:lem} we
have 
\[
M(\hat{X}_{\lambda_*},x) = x^2 \leq x^2 + (1+\sqrt{\beta})^2 = 
M(\hat{X}_r,x)\,.
\]
For $x\geq \beta^{1/4}$, by Lemma \ref{amse-tsvd:lem} and Theorem
\ref{inadmis:thm} we have 
\[
M(\hat{X}_{\lambda_*},x) \leq M(\hat{X}_r,x)\,.
\]
\qed

\subsubsection*{Proof of Theorems \ref{mmx-square:thm} and \ref{mmx-nonsquare:thm}}
Theorem \ref{mmx-square:thm} is a special case of Theorem
\ref{mmx-nonsquare:thm}.
By \eqref{amse:eq}, 
it is enough to consider the univariate function $x\mapsto
M(\hat{X}_\lambda,x)$ defined in \eqref{M:eq}. 
The theorem follows from Lemma \ref{amse:lem} using the following simple
observation.

  Let $0<\beta\leq 1$ and $\lambda > 1+\sqrt{\beta}$.
  Denote by $x_*(\lambda)$ the unique positive solution to the equation 
  $(x+1/x)(x+\beta/x) = \lambda^2$. Let 
  $\lambda_*$ be the unique solution to the equation in $\lambda$
  \[x_*^4(\lambda) - (\beta+1) x_*^2 - 3\beta=0\,.\]
  Then
  for the function $M(\hat{X}_\lambda,x)$ defined in \eqref{M:eq}, we have
  \begin{eqnarray*}
    \text{argmax}_{x>0} M(\hat{X}_\lambda,x) &=& x_*(\lambda)\\
    \text{argmin}_{\lambda> 1+\sqrt{\beta}}\max_{x>0}M(\hat{X}_\lambda,x) &=&
    \lambda_*\\
    \text{min}_{\lambda> 1+\sqrt{\beta}} \max_{x>0}M(\hat{X}_\lambda,x) &=&
    x_*(\lambda_*)^2\,.
  \end{eqnarray*}

\qed

Note that the least favorable situation occurs when 
$x_1=\ldots =x_\rk=x_*(\lambda)$, and that $x_*(\lambda)$ is precisely the
value of $x$ for which the corresponding limiting data singular value
satisfies $y_{n,i} \aslim \lambda$. In other words, the least
favorable situation occurs when the data singular values all coincide with each
other and with the chosen hard threshold.

\subsubsection*{Proof of Lemma \ref{sigmahat:lem}}
 Let $F_n$ denote the empirical cumulative distribution
function (CDF)  of the squared singular values of $Y_n$.
Write $y^2_{med,n} = Median(F_n)$ where $Median(\cdot)$ is a functional 
which takes as argument the CDF and delivers the median of that CDF.
Under our asymptotic framework,
almost surely, $F_n$ converges weakly to a limiting distribution, $F_{MP}$,
the CDF of the Mar\v{c}enko-Pastur distribution with shape parameter $\beta$
\cite{silverstein_book}.
This distribution has a positive density
throughout its support, in particular at its median.
The median functional is continuous for weak convergence
at $F_0$, and hence, almost surely,
\[
     y^2_{med,n} = Median(F_n) \aslim Median(F_0) = \mu_\beta, \qquad n
     \goto \infty\,.
\]
It follows that,
\[
\lim_{n\to\infty} \,\frac{\hat{\sigma}(Y_n)}{1/\sqrt{n}} \aseq
\lim_{n\to\infty} \,\frac{y_{med,n}}{\sqrt{\mu_\beta}} \aseq 1\,.
\]
\qed

\section{General white noise} \label{general_noise:sec}

Our results were formally stated for a sequence of models of the form
$Y=X+\sigma Z$, where $X$ is a non-random matrix to be estimated, and the entries
of $Z$ are i.i.d samples from a distribution that is orthogonally invariant (in
the sense that the matrix $Z$ follows the same distribution as $A Z B$, for
any orthogonal $A\in M_{m,m}$ and $B\in M_{n,n}$). While Gaussian noise is orthogonally
invariant, many common distributions, which one could consider to model white
observation noise, are not. 

One attractive feature of the discussion on optimal choice of singular value
hard threshold, presented above, is that the AMSE $\MSE(\hat{X},\V{x})$ only depends
on the signal matrix $X$ through its rank, or more specifically, through its 
nonzero singular values $\V{x}$. If the
distribution of $Z$ is not orthogonally invariant, MSE (or AMSE) losses this
 property and depends on properties of $X$ other than its rank. This point is
 discussed extensively in  \cite{Shabalin2010}.

In general white noise, which is not necessarily orthogonally invariant, one can
still allow MSE to depend on $X$ only through its singular values by 
placing a prior distribution on $X$ and shifting to a model where it is a
random, instead of a fixed, matrix.  Specifically, 
consider an alternative asymptotic framework to the one in Section
\ref{framework:subsec}, in which the sequence denoising problems
$Y_n=X_n+Z_n/\sqrt{n}$ satisfies the following assumptions:
\begin{enumerate}
  \item {\em General white noise:} The entries of $Z_n$ are i.i.d
    samples from a
    distribution with zero mean, unit variance and finite fourth moment.  
\item {\em Fixed signal column span and uniformly distributed signal singular vectors:}
Let the rank $\rk>0$ be fixed and choose a vector $\V{x}\in\R^\rk$ with
coordinates $\V{x}=(x_1,\ldots,x_\rk)$. Assume that for all $n$, 
\begin{eqnarray} \label{singvec:eq}
X_n = U_n \, diag(x_1,\ldots,x_r,0,\ldots,0) \, V_n'\,
\end{eqnarray}
is a singular value decomposition of $X_n$, where $U_n$ and $V_n$ are 
uniformly distributed random orthogonal matrices. Formally, $U_n$ and $V_n$ are
sampled from the Haar distribution on the
$m$-by-$m$ and $n$-by-$n$ orthogonal group, respectively.

  \item {\em Asymptotic aspect ratio $\beta$:}
    The sequence $\m_\n$ is such that $\m_\n / \n \to \beta$. 
 \end{enumerate}

 The second assumption above implies that $X_n$  a ``generic'' choice of matrix
 with nonzero singular values $\V{x}$, or equivalently, a generic choice of
 coordinate systems in which the linear operator corresponding to $X$ is
 expressed.

The results of \cite{Benaych-Georges2012}, which we have used, hold in this case
as well. It follows that Lemma \ref{amse:lem} and Lemma \ref{amse-tsvd:lem}, and
consequently all our main results, hold under this alternative framework. In
short, in general white noise, all our results hold if one is willing to only
specify the signal singular values, rather than the signal matrix, and consider
a ``generic'' signal matrix with these singular values.

\section{Empirical comparison of MSE with AMSE} \label{MSEvsAMSE:subsec}

\begin{figure*}[!t]
  \centering
    \includegraphics[width=7.7in, trim= 140 1 3 .5, clip=true]{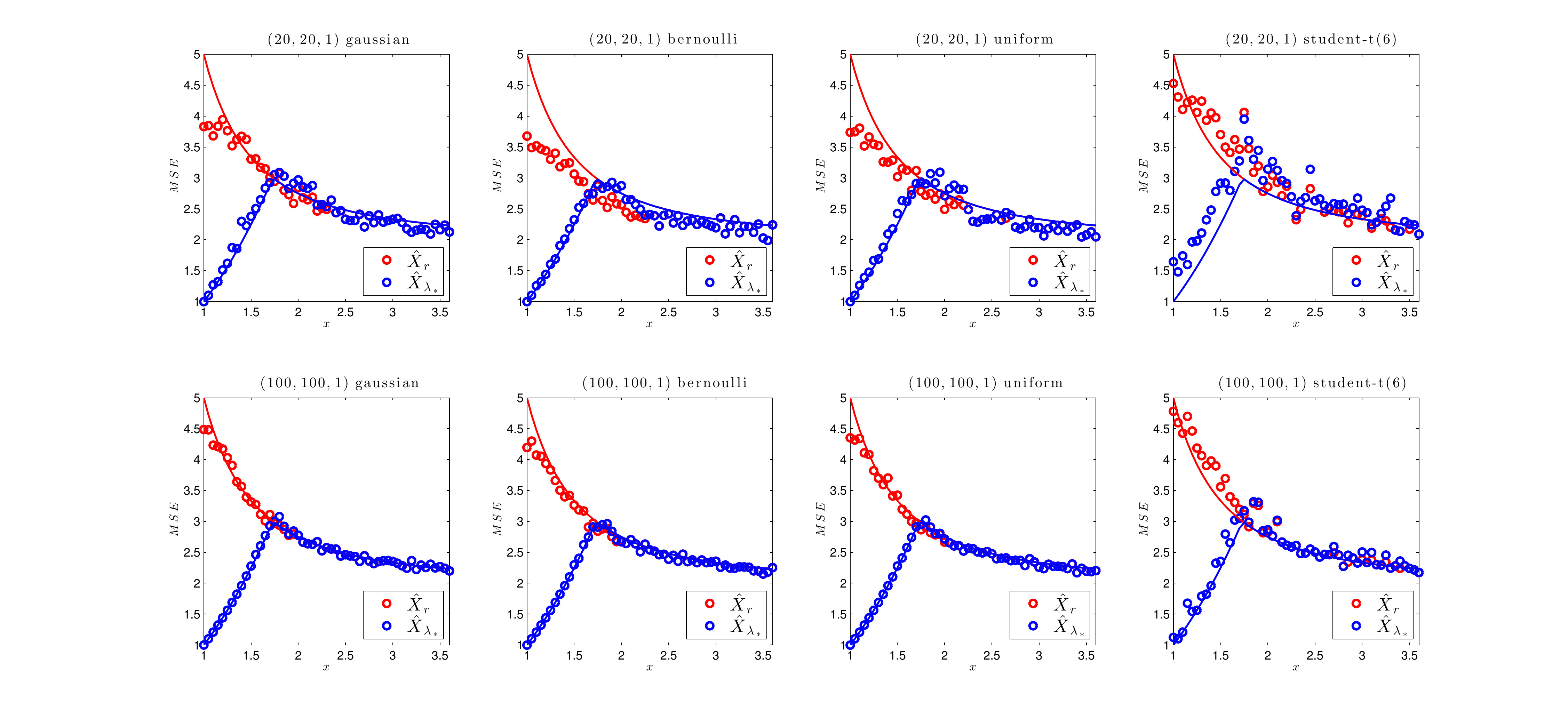}
    \includegraphics[width=7.7in, trim= 140 1 3 .5, clip=true]{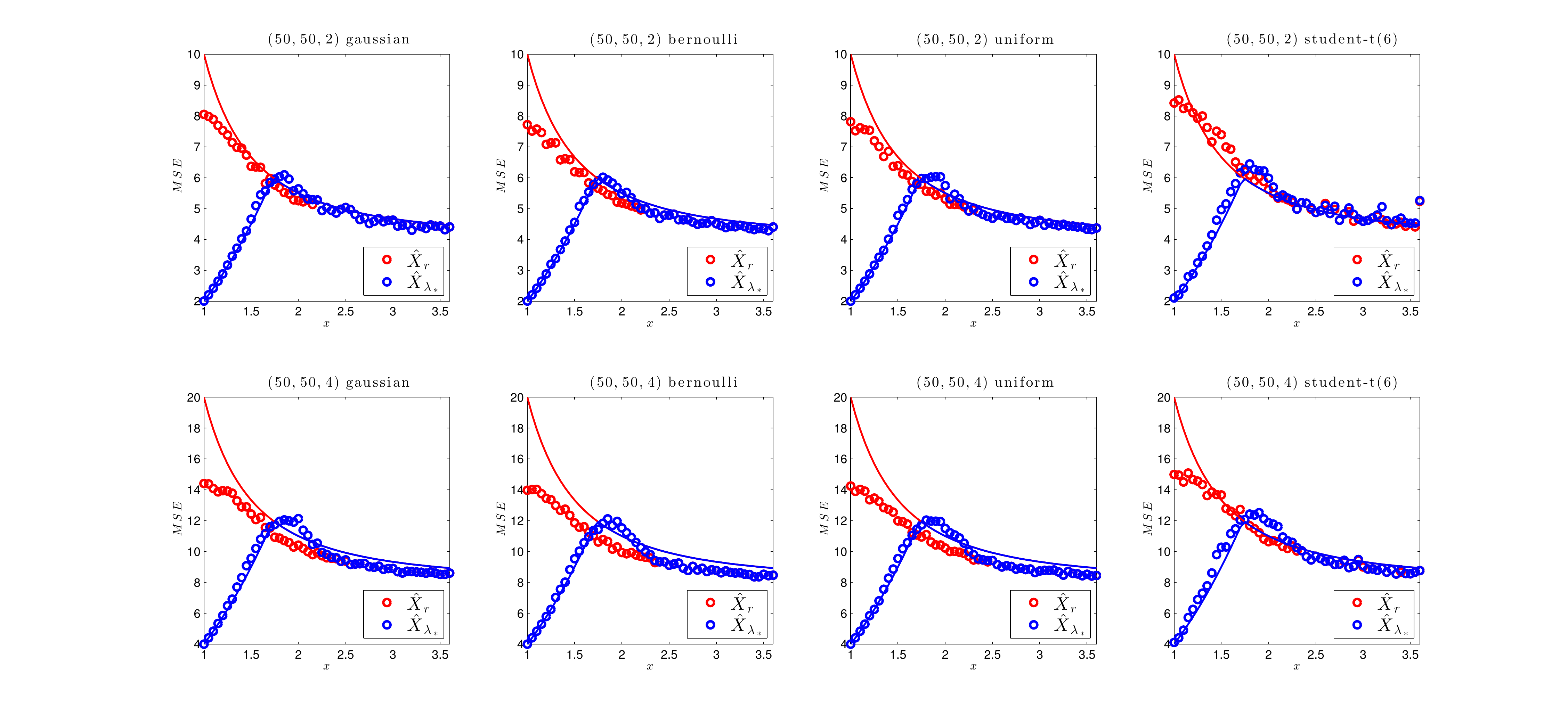}
  \caption{\scriptsize The AMSE (solid line) and empirical MSE (circles) 
  of TSVD $\hat{X}_\rk$ and
  optimal SVHT $\hat{X}_{\lambda_*}$ for $\beta=1$ and signal singular value
  $x\geq 1$ which correspond to leading data singular values that fall beyond
  the bulk egde.
  In a given panel, for a given value of $x$, the blue and the red dots were
  generated by first generating a signal matrix $X$, and then averaging each of
  the losses
  $\norm{\hat{X}(X+Z/\sqrt{n})-X}_F^2$ 
  over the same 50 Monte Carlo draws of the noise matrix $Z$.
  Each column of panels represents a
  different noise with zero mean and unit variance: Gaussian, Bernoulli on
  $\pm 1$, uniform on $[-0.5,0.5]$ and Student's $t$-distribution with $6$
  degreees of freedom;
 Panel titles indicate $(m,n,r)$ and the noise distribution.
 Top rows: $r=1$, different values of $m=n$. Bottom rows:  
 $m=n=50$, different values of $r$ (for $r>1$, signal singular values are all equal).
 {\em Reproducibility advisory:} script to
  generate figure, and to perform similar experiments, is included in code supplement \cite{code}.
 }
  \label{4a:fig}
\end{figure*}

We have calculated the exact
optimal threshold $\tau_*$ in a certain asymptotic framework. 
The practical significance of our results
hinges on the validity of the AMSE 
as an approximation to MSE, for values of $(m,n,r)$ and error distributions
encountered in practice. This in turn depends on the simultaneous convergence of
three terms: 
\begin{itemize}
  \item Convergence of the top data singular values $y_{n,i}$ ($1\leq i \leq
    r$) to the limit in Lemma \ref{y-asy:lem},
  \item Convergence of the angle between the top data singular vectors $\V{u}_{n,i},\V{v}_{n,i}$
    and their respective signal singular vectors 
    to the limit in Lemma \ref{inner-asy:lem}, and
  \item Convergence of the rest of the data singular values $y_{n,i}$ ($r+1\leq i
    \leq m)$ to the interval $[0,1+\sqrt{\beta}]$.
\end{itemize}
Analysis of each of these terms for specific error distributions is beyond our
current scope. Figure \ref{4a:fig} contains a few sample comparisons of AMSE and
empirical MSE we have performed. The matrix sizes and number of Monte Carlo
draws are small enough to demonstrate that AMSE is a reasonable approximation
even for relatively small low-rank matrices.  As convergence of the empirical
spectrum to its limit is known to depend on moments of the underlying
distributions, we include results for different error distributions.  AMSE is
found to be a useful proxy to MSE even in small matrix sizes.  AMSE of SVHT was
found to be inaccurate when: (i)  the rank fraction is nontrivial (e.g $n=50$,
$r=4$ shown at the bottom of Figure \ref{4a:fig}); (ii) the threshold $\lambda$
is very close to the approximate bulk edge $1+\sqrt{m/n}$. In case (i),
interaction effects between singular values, which are ignored in our asymptotic
framework, start to have non-negligible effect. In case (ii), where the
discontinuity of the SVHT nonlinearity is placed close to the bulk edge, the
distribution of the largest ``non-signal'' singular value $y_{n,r+1}$, which is
known in some cases to be asymptotically a Tracy-Widom distribution
\cite{Johnstone2001}, becomes important. Indeed, some data singular values from
the bulk manage to pass the threshold $\lambda$ and cause their singular vectors
to be included in the estimator $\hat{X}_\lambda$. Our derivation of AMSE
assumed however that no such singular vectors are included in $\hat{X}_\lambda$,
since  $y_{n,r+1}\aslim 1+\sqrt{\beta}$.  Note however that the main
recommendation of this paper is that one should {\em not} threshold at or near
he bulk edge, as explained in detail above. Therefore, from a practical
perspective, the inaccuracy of AMSE for SVHT with $\lambda$ near the bulk edge
is slightly irrelevant.

\section{Conclusion} \label{conclusion:sec}

The asymptotic framework considered here is perhaps the simplest nontrivial
model for matrix denoising. It allows one to calculate, in AMSE, basically any
quantity of interest, for any denoiser of interest.  The fundamental elements of
matrix denoising in white noise, which underly more complicated models, are
present yet understandable and quantifiable. For example, the AMSE of any
denoiser based on singular value shrinkage contains a component due to noise
contamination in the data singular vectors, and this component determines a
fundamental lower bound on AMSE.

We conjecture that results calculated in this model, which are not attached to a
specific assumption on rank (e.g, the constants in Table
\ref{mmx-square-l1:tab}, which determine the minimax AMSE over nuclear norm
balls) remain essentially correct in more complicated models.  

The decision-theoretic landscape as it appears through the naive prism of our
asymptotic framework is extremely simple: there is a unique admissible hard
thresholding rule, and moreover a unique admissible shrinkage rule, for singular
values. This is of course quite different from the situation encountered, for
example, in estimating normal means. The reason is the extreme simplicity of our
model. For example, we have replaced the data singular values, which are random for finite
matrix size, with their almost sure limits, and in effect neglected their random
fluctuations around these limits. These fluctuations are now well understood (see for example
\cite{Bai2008,Shi2013}). We have ignored this structure. However, including these second-order terms in
the asymptotic distributions is only likely to achieve second-order improvements
in MSE over our suggested optimal truncation threshold.

\section*{Reproducible Research}

In the code supplement \cite{code} we offer a Matlab software library that
includes:
\begin{enumerate}
  \item A function that calculates the optimal shrinkage coefficient in known or
    unknown noise level.
  \item Scripts that generate each of the figures in this paper.
  \item A script that generates figures similar to Figure \ref{4a:fig}, which
    compare AMSE to MSE in various situations.
\end{enumerate}

\section*{Acknowledgements}

The authors would like to thank Andrea Montanari for pointing  to the work of
Shabalin and Nobel, Drew Nobel and Sourav Chatterjee for helpful discussions,
Art Owen for pointing to the work of Perry,
and the anonymous referees for their useful suggestions.  This work was
partially supported by NSF DMS 0906812 (ARRA). MG was partially supported by a
William R. and Sara Hart Kimball Stanford Graduate Fellowship.

%

\begin{IEEEbiography}[{\includegraphics[width=1in,height=1.25in,clip,keepaspectratio]{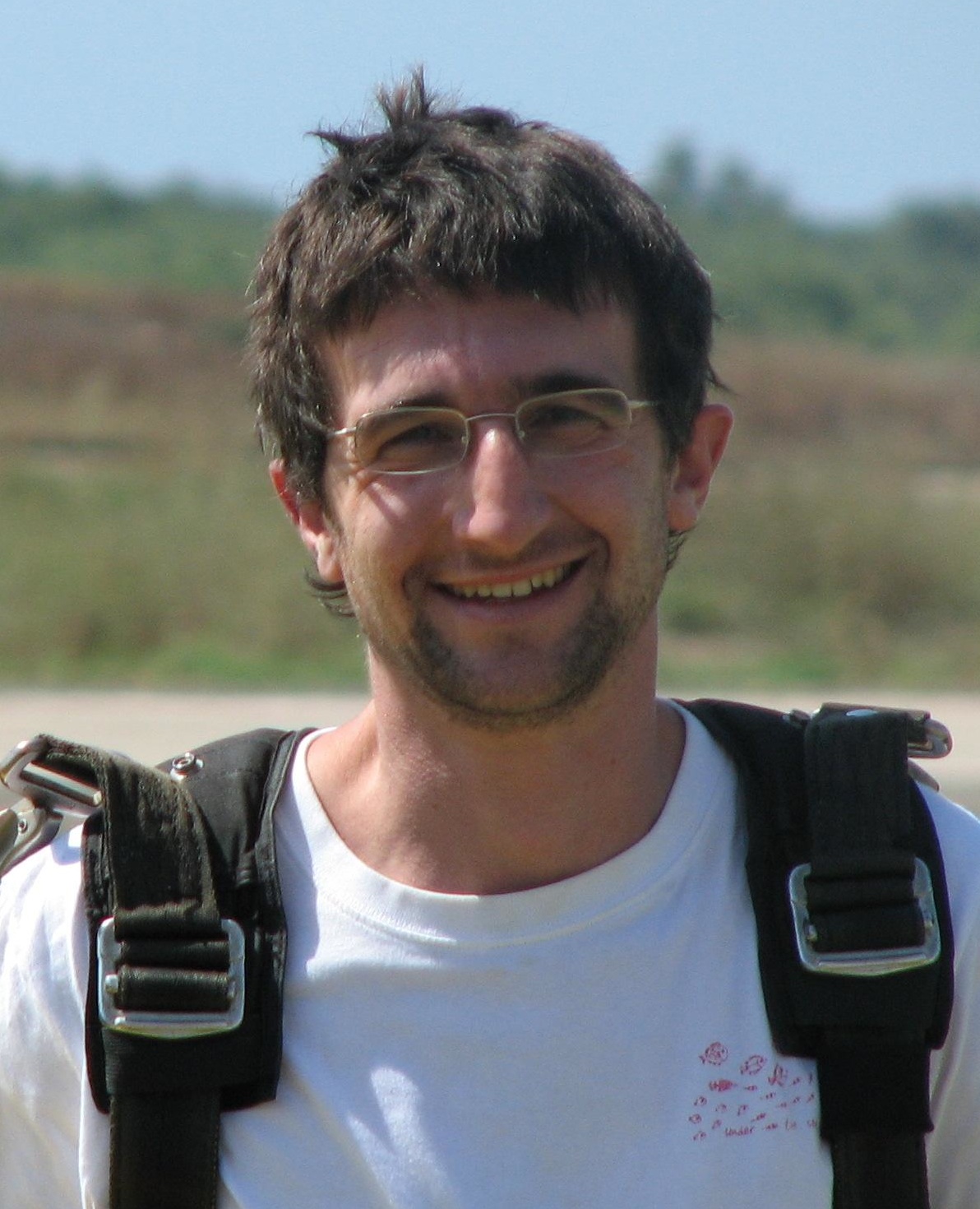}}]{Matan
  Gavish} received the dual B.Sc. degree in Mathematics and Physics from Tel
  Aviv University (TAU) in 2006 and the M.Sc. degree in Mathematics from the
  Hebrew University of Jerusalem in 2008. He is currently a doctoral student in
  Statistics at Stanford University, in collaboration with the Yale University
  program in Applied Mathematics. His research interests include applied
  harmonic analysis, high-dimensional statistics and computing.  He was in the
  Adi Lautman Interdisciplinary Program for outstanding students at TAU from
  2002 to 2006 and held a William R. and Sara Hart Kimball Stanford Graduate
  Fellowship from 2009 to 2012.  
\end{IEEEbiography}

\begin{IEEEbiography}[{\includegraphics[width=1in,height=1.25in,clip,keepaspectratio]{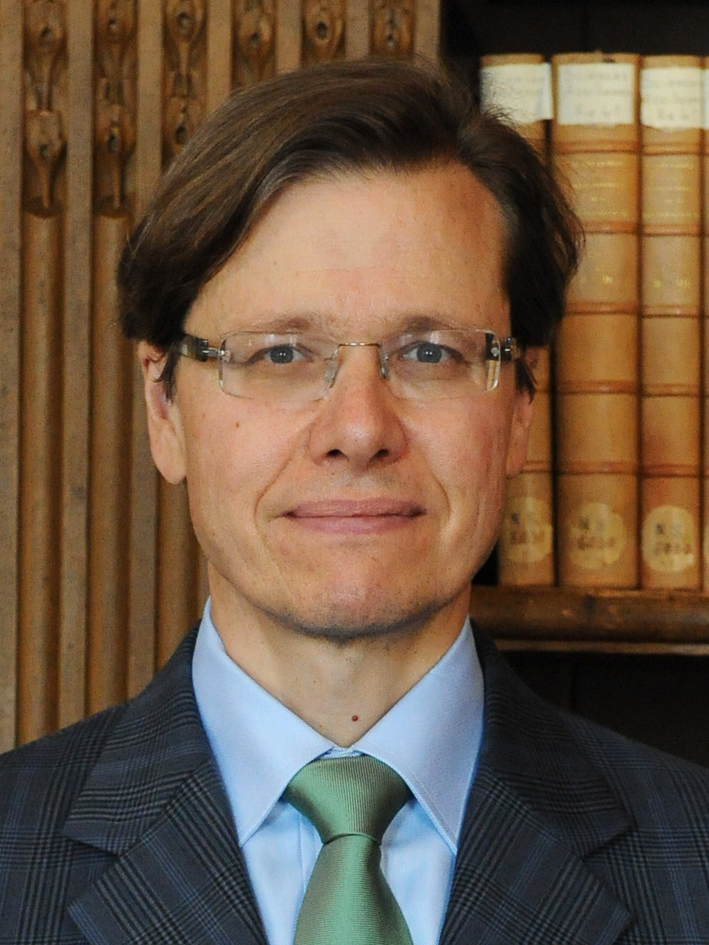}}]{David
  L. Donoho} is a professor at Stanford University.
His research interests include computational
harmonic analysis, high-dimensional geometry, and
mathematical statistics. Dr. Donoho received the
Ph.D. degree in Statistics from Harvard University,
and holds honorary degrees from University of Chicago
and Ecole Polytechnique Federale de Lausanne.
He is a member of the American Academy of Arts and
Sciences and the US National Academy of Sciences, and a foreign associate
of the French Acad\'emie des sciences.  
\end{IEEEbiography}


\vfill
\vfill

\end{document}